\documentclass[sigconf,authorversion,nonacm]{acmart}
 \pdfoutput=1
\AtBeginDocument{%
  }

\setcopyright{none}

\usepackage{color}
\usepackage{booktabs}
\usepackage{url}
\usepackage{tabularx}
\usepackage{bm}
\usepackage{mdframed}
\usepackage{enumitem}
\usepackage{algorithm}
\usepackage{algpseudocode}
\usepackage{amsmath}
\usepackage{scalerel}
\usepackage{tablefootnote}
\usepackage{enumitem}
\usepackage{tablefootnote}
\usepackage[export]{adjustbox}
\usepackage[normalem]{ulem}
\usepackage{color, colortbl}
\usepackage{multirow}
\usepackage{subcaption}

\definecolor{Gray}{gray}{0.9}

\definecolor{burgundy}{rgb}{0.5, 0.0, 0.13}
\definecolor{cadmiumgreen}{rgb}{0.0, 0.42, 0.24}
\definecolor{burntorange}{rgb}{0.8, 0.33, 0.0}
\definecolor{bluegray}{rgb}{0.4, 0.6, 0.8}
\definecolor{emerald}{rgb}{0.31, 0.78, 0.47}
\definecolor{applegreen}{rgb}{0.55, 0.71, 0.0}
\definecolor{ao(english)}{rgb}{0.0, 0.5, 0.0}

\newcommand{\malicious}[1]{\textcolor{purple}{#1}}

\mathchardef\mhyphen="2D 

\newcommand{\ours}{\textsc{LiSA}}

\newcommand{\etal}{\textit{et al.}}

\newcommand{\NN}{\mathbb{N}}

\newcommand{\XX}{\mathbb{X}}

\newcommand{\size}[1]{\lvert #1 \rvert}

\newcommand{\TRUE}{\textsc{true}}
\newcommand{\FALSE}{\textsc{false}}
\newcommand{\asim}{\mathcal{O}}

\newcommand{\users}{\mathcal{U}}
\newcommand{\dusers}{\mathcal{D}}
\newcommand{\cusers}{\mathcal{C}}
\newcommand{\committee}{\mathcal{K}}
\newcommand{\backupset}{\mathcal{L}}

\newcommand{\prob}[1]{\Pr \left[ #1 \right]}

\newcommand{\ExpVal}[2]{\mathbb{E}_{#1}[#2]}

\newcommand{\hypergeom}{\mathcal{HG}}

\newcommand{\sep}{\lambda}

\newcommand{\sk}{sk} 
\newcommand{\pk}{pk} 
\newcommand{\ct}{c} 

\newcommand{\C}{\mathcal{C}} 

\newcommand{\Gen}{\mathsf{Gen}} 
\newcommand{\Recon}{\mathsf{Rec}} 

\newcommand{\nchoosek}{\mathsf{Select}}

\renewcommand{\SS}{\mathsf{SS}}
\newcommand{\SSshare}{\SS.\mathsf{Share}}
\newcommand{\SSrecon}{\SS.\Recon}

\newcommand{\KA}{\mathsf{KA}}

\newcommand{\KAgen}{\KA.\Gen}

\newcommand{\KAenc}{\KA} 
\newcommand{\KAprf}{\KA} 
\newcommand{\ssk}{SK}
\newcommand{\ppk}{PK}

\renewcommand{\AE}{\mathsf{AE}}
\newcommand{\AEenc}{\AE.\mathsf{Enc}}
\newcommand{\AEdec}{\AE.\mathsf{Dec}}

\def\CommGen{{\sf COMM.Gen }}
\def\CommVfy{{\sf COMM.Vfy }}
\def\CommEqv{{\sf COMM.Eqv }}
\def\CommExt{{\sf COMM.Ext }}

\def\commi{{\sf comm}}

\newcommand{\Ssk}{sk^{s}} 
\newcommand{\Spk}{pk^{s}} 
\newcommand{\DS}{\mathsf{DS}}
\newcommand{\DSgen}{\DS.\mathsf{Gen}}
\newcommand{\DSsign}{\DS.\mathsf{Sign}}
\newcommand{\DSverif}{\DS.\mathsf{Verif}}

\newcommand{\funct}{\mathcal{F}}
\newcommand{\functSUM}{\funct^{SUM}}
\newcommand{\functSUMVERIF}{\funct^{SecVerAgg}}
\newcommand{\adv}{\mathnormal{A}}

\newcommand{\REAL}{\mathrm{REAL}}
\newcommand{\SIM}{\mathrm{SIM}}

\newcommand{\RO}{\mathcal{H}}

\newtheorem{theorem}{Theorem}

\begin{document}

\title{\ours: LIghtweight single-server Secure Aggregation\\ with a public source of randomness}

\author{Elina van Kempen}
\email{evankemp@uci.edu}
\affiliation{%
  \institution{UC Irvine}
  \country{United States}
}
\author{Qifei Li}
\email{qifei.li@neclab.eu}
\affiliation{%
  \institution{NEC Laboratories Europe}
  \country{Germany}
}

\author{Giorgia Azzurra Marson}
\email{giorgia.marson@neclab.eu}
\affiliation{%
  \institution{NEC Laboratories Europe}
  \country{Germany}
  }

\author{Claudio Soriente}
\email{claudio.soriente@neclab.eu}
\affiliation{%
  \institution{NEC Laboratories Europe}
  \country{Germany}
}


\begin{abstract}
  Secure Aggregation (SA) is a key component of privacy-friendly federated learning applications, where the server learns the sum of many user-supplied gradients, while individual gradients are kept private. State-of-the-art SA protocols protect individual inputs with zero-sum random shares that are distributed across users, have a per-user overhead that is logarithmic in the number of users, and take more than 5 rounds of interaction.

  In this paper, we introduce {\ours}, an SA protocol that leverages a source of public randomness to minimize per-user overhead and the number of rounds. In particular, {\ours} requires only two rounds and has a communication overhead that is asymptotically equal to that of a non-private protocol---one where inputs are provided to the server in the clear---for most of the users. In a nutshell, {\ours} uses public randomness to select a subset of the users---a committee---that aid the server to recover the aggregated input. Users blind their individual contributions with randomness shared with each of the committee members; each committee member provides the server with an aggregate of the randomness shared with each user. Hence, as long as one committee member is honest, the server cannot learn individual inputs but only the sum of threshold-many inputs.
  We compare {\ours} with state-of-the-art SA protocols both theoretically and by means of simulations and present results of our experiments. We also integrate {\ours} in a Federated Learning pipeline and compare its performance with a non-private protocol.
\end{abstract}



\maketitle

\section{Introduction}

Secure Aggregation (SA) allows a set of parties to compute a linear function (e.g., the sum) of their inputs while keeping each party's input private. Recently, SA protocols have been proposed as a privacy-preserving mechanism for federated learning , where a large number of users and a central server train a joint model by leveraging user-private gradients~\cite{DBLP:conf/ccs/BonawitzIKMMPRS17,DBLP:conf/ccs/BellBGL020}.

State-of-the-art SA requires users to provide noisy version of their inputs to the server, so that when all noisy inputs are aggregated, the noise cancels out and the server learns the aggregated gradient.
An ideal SA protocol should minimize the overhead at the user and the number of rounds, since in many FL applications, users are devices with limited computing power and an erratic online behavior. This is typically the case when FL is employed to train voice recognition or text prediction models on mobile phones~\cite{DBLP:conf/mlsys/BonawitzEGHIIKK19}.

Existing SA protocols offer security against static corruptions of the server and a fraction of the users, and are robust to user drop-outs throughout the protocol execution. Among available protocols for SA in federated learning, the one proposed by Bell~\etal~\cite{DBLP:conf/ccs/BellBGL020} has communication overhead of $\mathcal{O}(\log^2 n+m)$ for users and $\mathcal{O}(n(\log^2 n+m))$ for the server, given~$n$ users, each with an input of size $m$. Further, the protocol in~\cite{DBLP:conf/ccs/BellBGL020} requires 5.5 rounds of interaction in case of malicious adversaries. Note that a non-private aggregation protocol has a communication overhead of $\mathcal{O}(m)$ for a user, $\mathcal{O}(nm)$ for the server, and requires a single round.

\paragraph*{Our contribution}
We study how to design efficient and secure aggregation protocols for federated learning, assuming the availability of a random beacon service, i.e., a source of public randomness.
Random beacons can be instantiated with financial data~\cite{DBLP:conf/uss/ClarkH10} or crypto-currencies~\cite{DBLP:journals/iacr/BonneauCG15}. Alternatively, one can use a trusted source of randomness (e.g., \url{https://beacon.nist.gov/home}) or distributed protocols to generate public randomness~\cite{DBLP:conf/sp/SytaJKGGKFF17,DBLP:conf/acns/CascudoD17,DBLP:conf/sp/DasKI022}.
Previous work that leverages random beacon services include secure messaging~\cite{DBLP:conf/nsdi/KwonLD20,DBLP:conf/sosp/KwonCDF17,DBLP:conf/sosp/TyagiGLZZ17}, sortition~\cite{DBLP:conf/sosp/GiladHMVZ17} or e-voting~\cite{DBLP:conf/uss/Adida08}.

In this paper we present \emph{LIghtweight Secure Aggregation} ({\ours}), a secure aggregation protocol that leverages a public random beacon to minimize overhead.
In particular, {\ours} protects user inputs with random noise. Thus, we use the random beacon at each learning epoch to select a small subset of the users, a \emph{committee}, that hold the seeds necessary to remove the noise from the inputs provided by the users. The technique we employ to add and remove noise from the inputs ensures that, as long as one committee member is honest, a compromised server cannot learn individual user inputs but only the sum of threshold-many inputs. Intuitively, the random choice of committee members guarantees that at least one committee member is honest (with high probability), despite a fraction of the users may be compromised.

By using a randomly-selected committee, {\ours} does not require each user to secret-share the randomness used to add noise to her input with  other peers like in~\cite{DBLP:conf/ccs/BonawitzIKMMPRS17,DBLP:conf/ccs/BellBGL020}. Users of {\ours} add noise to their inputs with randomness derived from a non-interactive key agreement with each of the committee members. Hence, committee members must provide the server with the same randomness so that inputs can be de-noised. In order to protect individual user inputs, honest committee members do not share with the server the noise used to blind an individual input, but rather provide the server with aggregated noise, so that the server can de-noise only aggregated inputs.

{\ours} exhibits a communication overhead of $\mathcal{O}(m)$ for most of the users and $\mathcal{O}(nm)$ for the server. Further, most users in {\ours} are required to be online for only~2 rounds. 
We theoretically compare {\ours} with previous work in terms of overhead. Further, we use a simulator to compare the per-user communication overhead of our protocol with the one of~\cite{DBLP:conf/ccs/BellBGL020} and show that {\ours} has a lower average number of messages per user. 

Finally, we show how to protect integrity of the aggregated inputs by means of cryptographic commitments.

The rest of the paper is organized as follows.
Section~\ref{sec:related} surveys related work while Section~\ref{sec:background} provides background on relevant cryptographic primitives. We define the problem of secure aggregation and present our protocol in Section~\ref{sec:lisa}. Section~\ref{sec:analysis} provides correctness and security arguments. Section~\ref{sec:performance} shows theoretical overhead, introduces our prototype, and presents the results of our evaluation. Section~\ref{sec:discussion} discusses how to add integrity protection to {\ours} and how to choose relevant system parameters. We provide concluding remarks in Section~\ref{sec:conclusions}.

\section{Related work}
\label{sec:related}
\begin{table*}[t]
\noindent\scalebox{0.7}{
    \begin{tabular}{lccccccp{2cm}}
    \toprule
    & \multicolumn{2}{c}{\textbf{Communication}} & \multicolumn{2}{c}{\textbf{Computation}} & \textbf{{\#} rounds} & \textbf{Corruption/dropout thresholds} & \textbf{Requirements}\\
       & User & Server & User & Server &  &  & \\
    \midrule
    Bonawitz~\etal~\cite{DBLP:conf/ccs/BonawitzIKMMPRS17}  & $\asim(n+m)$              & $\asim(n^2+nm)$ &  $\asim(n^2+nm)$ & $\asim(mm^2)$ &  5  & $\delta<1/3$, $\gamma<1/3$ & PKI \\
    Bell~\etal~\cite{DBLP:conf/ccs/BellBGL020}             & $\asim(\log^2 n+m)$       & $\asim(n\log^2 n+nm)$ &  $\asim(\log^2 n+m\log n)$ & $\asim(n\log^2 n+nm\log n)$ & 5.5 & $2\delta+\gamma<1$ & PKI \\
    Stevens~\etal~\cite{DBLP:conf/uss/StevensSVRCN22}      & $\asim(n+m)$              & $\asim(nm)$ &  $\asim(m+n\log n)$ & $\asim(nm+n\log n)$ &  3  & honest majority & none \\
    MicroFedML \cite{DBLP:journals/iacr/GuoPSBB22}         & $\asim(m\log n)$          & $\asim(nm\log n)$ &  $\asim(nm)$ & $\asim(nm)$ &  3  & $2\delta+\gamma<1$ & PKI \\
    \midrule
    \multirow{2}{*}{Flamingo \cite{cryptoeprint:2023/486}} & Regular: $\asim(m+A)$     & \multirow{2}{*}{$\asim(n(m+A+L))$} &  Regular: $\asim(Am+L^2+n\log n)$    & \multirow{2}{*}{$\asim(nm+n^2)$} &  1 & \multirow{2}{*}{$2\delta _D + \gamma _D <1/3$} & \multirow{2}{*}{\parbox{2cm}{PKI, trusted randomness}}\\
    & Decryptors: $\asim(L+An)$ &                    &  Decryptors: $\asim(L+n)$ &                    &  3  &  &    \\
    \midrule
    \multirow{3}{*}{{\ours} (this work)}                   & Regular: $O(m)$           & \multirow{3}{*}{$\asim(nm)$} & Regular: $\asim(m)$    & \multirow{3}{*}{$\asim(nm)$} &  2  & \multirow{3}{*}{$2\delta+\gamma<1$} & \multirow{3}{*}{\parbox{2cm}{PKI, trusted randomness}}\\
    & Committee: $\asim(n+m)$              &                    & Committee: $\asim(nm)$ &                     &  3  &                                                          \\
    & Backups:   $\asim(1)$              &                    & Backup: $\asim(1)$    &                     &  5  &                                                           \\
    \midrule
    \midrule
    Non-private protocol & $\asim(m)$  & $\asim(nm)$ & $\asim(1)$ & $\asim(nm)$ & 1 & N/A & N/A   \\
    \bottomrule
    \end{tabular}
    }
    \caption{Comparison between prior secure aggregation protocols with (malicious security) and~\ours.
    We use $n$ to denote the number of users and~$m$ to denote the size of the user's input.
    Flamingo considers a set of decryptors of size~$L$; moreover, users distribute their masks with a set of neighbors of size~$A$ (similarly to Bell~\etal). According to their analysis, $L$ is $\asim(1)$ and $A$ is $\asim(\log n)$. Also, $\delta _D$ denotes the fraction of decryptors that drop-out while $\gamma_D$ denotes the fraction of decryptors that are honest.
    }
    \label{table:related}
  \end{table*}

In the context of federated learning, single-server secure aggregation was introduced by Bonawitz~\etal~\cite{DBLP:conf/ccs/BonawitzIKMMPRS17}. Their protocol requires each user to blind her input with two blindings, one private and one shared with another peer. Both blindings are secret shared with all other users and a reconstruction protocol allows the server to remove the blindings from the sum of individual blinded inputs. As such, the protocol in~\cite{DBLP:conf/ccs/BonawitzIKMMPRS17} has an overhead that is linear in the number of users. Further, it allows some users to drop-out and withstands collusion between a malicious server and a fraction of the users. To the best of our knowledge, only four other single-server secure aggregation protocols consider malicious behaviour of corrupted parties. Those are the protocols by Bell~\etal~\cite{DBLP:conf/ccs/BellBGL020}, the one proposed by Stevens~\etal~\cite{DBLP:conf/uss/StevensSVRCN22}, MicroFedML by Guo~\etal~\cite{DBLP:journals/iacr/GuoPSBB22}, and Flamingo by Ma~\etal~\cite{cryptoeprint:2023/486}.

Bell~\etal~\cite{DBLP:conf/ccs/BellBGL020} propose an extension of~\cite{DBLP:conf/ccs/BonawitzIKMMPRS17} that reduces the communication overhead by partitioning users in groups of $\log n$ size and  restricting communication among members of the same group.
Stevens~\etal\cite{DBLP:conf/uss/StevensSVRCN22} use differential privacy to provide secure aggregation but their communication overhead is not on par with~\cite{DBLP:conf/ccs/BellBGL020}.
MicroFedML~\cite{DBLP:journals/iacr/GuoPSBB22} carries out aggregation ``at the exponent'' of a cyclic group and requires discrete-log computation. Hence, MicroFedML is only suitable when the bit-length of the sum of the inputs  $\ell$ is ``small'' (e.g., 20 bits)---hence, the input of a single users must be smaller than $2^\ell/n$. Differently, \ours{} can work with large-domain inputs. MicroFedML has a setup phase of three rounds, and aggregation takes three rounds for all users. Storage requirement due to setup is $\asim(n)$. Aggregation in \ours{} takes 2 rounds for most users and has no setup phase. If we added a setup phase to {\ours} where users exchange keys similar to MicroFedML, then aggregation in {\ours} would take only one round for most users, but storage requirement would become $\asim(n)$.
An independent and concurrent work by Ma~\etal~\cite{cryptoeprint:2023/486} introduces Flamingo, a secure aggregation protocol that, similarly to~\ours, leverages a trusted source of randomness to select a committee (called ``decryptors'' in~\cite{cryptoeprint:2023/486}) to aid the server in computing the aggregation from the masked inputs submitted by users. We now list the main differences between Flamingo and \ours{}. First, Flamingo works by means of linearly homomorphic encryption whereas \ours{} leverages zero-sum random shares.  Users in Flamingo must be active for one round whereas users in \ours{} stay online for two rounds.
Communication for regular users is $\asim(m+A)$ in Flamingo---where $A$ is the size of the neighborhood of a user and should be $\asim(\log n)$---while \ours{} has communication complexity $\asim(m)$ for regular users. Flamingo requires a setup phase of $7$ rounds, and a periodic share-transfer phase where a new set of decryptors is selected; the share-transfer phase takes $4$ rounds so that the secret key shares are transferred from the current set of decryptors to the new set of decryptors. Changing the set of committee members in \ours{} requires no communication among protocol participants. Finally, Flamingo tolerates less corruptions and drop-outs compared to \ours. In particular, during the setup phase of Flamingo, at least $\frac{2}{3}$ of the decryptors must be honest and alive; differently, \ours{} remains secure as long as a single committee member is honest and alive.

Other secure aggregation protocols can only withstand a semi-honest adversary and, at times, they do not allow the server to collude with users. Choi~\etal~\cite{DBLP:journals/corr/abs-2012-05433} exploit the same idea as ~\cite{DBLP:conf/ccs/BellBGL020}, and partition users to form Erdős-Rényi graphs.
Turbo-Aggregate \cite{DBLP:journals/jsait/SoGA21a} and SwiftAgg+ \cite{DBLP:journals/corr/abs-2203-13060} also take advantage of dividing users into groups, but require certain users to have a direct communication link with each other. Turbo-Aggregate needs a non-constant number of communication rounds to complete the protocol, which may increase the likeliness of user drop-out.
FastSecAgg~\cite{DBLP:journals/corr/abs-2009-11248} uses a novel additive secret sharing scheme based on Fast Fourier Transform. While faster secret sharing speeds are achieved, the communication overhead for user is still linear. 

The goal of SparseSecAgg \cite{DBLP:journals/corr/abs-2112-12872} and HeteroSAg \cite{DBLP:journals/tcom/ElkordyA22} is to make secure aggregation possible in the presence of users that have different resources. In SparseSecAgg, users do not need to send the entire model for aggregation, gradient sparsification is possible. With the availability of heterogeneous quantization in HeteroSAg, users with different communication resources can join the federated learning application. Both protocols exhibit linear communication overhead in terms of number of users. Truex~\etal~ \cite{DBLP:conf/ccs/TruexBASLZZ19} leverages differential privacy, only works for semi-honest adversaries, and makes no mention of possible user drop-outs. Ma~\etal~ \cite{DBLP:journals/ijis/MaNSL22} present xMK-CKKS, a multi-key homomorphic encryption protocol, that allows for secure aggregation with a semi-honest server. Fereidooni~\etal~ \cite{DBLP:conf/sp/FereidooniMMMMN21} propose SAFELearn, a generic design of secure aggregation for federated learning, where either Fully Homomorphic Encryption (FHE) or Multi-Party Computation (MPC) can be used. If MPC is used, at least two non-colluding servers are needed. The communication costs in case FHE is used were not reported. The protocol is secure against a semi-honest server, no information is given on the users being fully honest or semi-honest. Also, it is not clear if user-server collusion is allowed.

We summarize related work and compare it to {\ours} in Table~\ref{table:related}.
The table also reports overhead and number of rounds of a non-private protocol where users sends their inputs to the server in the clear.

In July 2023 we became aware of an independent and concurrent work~\cite{DBLP:journals/corr/abs-2304-03841} that introduces \textit{e-SeaFL}, an SA protocol that shares some basic ideas with {\ours}.
Namely, e-SeaFL leverages a small set of parties, dubbed ``assisting nodes'', to help the server in computing the aggregation. 
The assisting nodes in e-SeaFL are similar in spirit to the committee members in {\ours}, and both protocols retain privacy as long as one of such nodes is honest.

However, e-SeaFL makes stronger trust assumptions on the assisting nodes compared to {\ours}.
Firstly, the set of assisting nodes is known upfront and stays unchanged.
Although the manuscript suggests that a rotating set of users could serve as assisting nodes, as the protocol relies on a one-time setup (similarly to MicroFedML~\cite{DBLP:journals/iacr/GuoPSBB22} and Flamingo\cite{cryptoeprint:2023/486}), refreshing the assisting nodes would require a new setup each time.
Moreover, correctness of e-SeaFL (implicitly) requires that all assisting nodes remain online for the entire duration of the training process. Instead, {\ours} can tolerate dropouts in the committee by employing a backup mechanism to recover the shared secrets of offline committee members.

Finally, the manuscript describes an extension of e-SeaFL that should guarantee aggregation integrity. However, this extension requires one of the assisting nodes to generate a system-wide secret~$\rho$ and to securely distribute it to the all other nodes in the system (both assisting nodes and regular nodes). As the integrity definition relies on the confidentiality $\rho$, e-SeaFL can guarantee integrity only if all nodes are trusted---that is, if the adversary corrupts one node (be it an assisting node or a regular one), integrity can be broken. Differently \ours+ (cf. Figure~\ref{fig:protocol:integrity}) guarantees integrity despite an adversary that corrupts the server and a number of nodes.

\section{Background}
\label{sec:background}

\paragraph{Notations}
For $m,n\in\NN$, we denote by~$[m..n]$ the set of integers $\{i : m\leq i \leq n\}$, and we use the shorthand~$[n]:=[1..n]$.
We write $x\gets v$ for the assignment of value~$v$ to variable~$x$.

\subsection{Cryptographic Primitives}
\label{sec:background:crypto:primitives}
We now briefly describe the cryptographic primitives used in \ours. We omit their complete definitions because they have been used (and defined) in previous work on secure aggregation~\cite{DBLP:conf/ccs/BonawitzIKMMPRS17,DBLP:conf/ccs/BellBGL020}.

A secret-sharing scheme $\SS$ with algorithms $\SSshare$ and $\SSrecon$. The sharing algorithm $\{s_i\}_{i\in [n]}
\gets\SSshare(s,t,n)$ takes as input a secret $s$, a reconstruction threshold $t$ and a target number of shares $n$ and outputs a set of shares $\{s_i\}_{i\in[n]}$.
The reconstruction algorithm $s\gets\SSrecon(s_{i_1},\dots,s_{i_l})$ takes as input a set of shares
$s_{i_1},\dots,s_{i_l}$ such that $l\geq t$ and outputs a secret $s$. In practice, $\SS$ can be instantiated with Shamir secret-sharing~\cite{DBLP:journals/cacm/Shamir79}.

A non-interactive key agreement scheme $\KA$ with algorithm $(\sk_i,\pk_i)\gets\KAgen(1^\sep)$ to generate a key-pair and algorithm $k_{i,j}\gets\KAprf(\sk_i,\pk_j)$ to generate a shared key between party $i$ and party $j$.
We add a ``context'' string $\textsc{ctx}$ to the key agreement protocol and write $\KA(\textsc{"ctx"};\sk_i,\pk_j)$ so to allow party $i$ and party $j$ to derive multiple keys, by using different context strings. $\KA$ can be instantiated with DHKE~\cite{DBLP:journals/tit/DiffieH76} followed by a key-derivation function.

A pseudorandom number generator (PRG)~$F$ that can be instantiated with AES-CTR~\cite{DBLP:conf/ccs/BonawitzIKMMPRS17}.

An authenticated encryption scheme $\AE$. 
We denote by $\ct\gets\AEenc(k,m)$ the encryption of message $m$ under key $k$, and $m\gets\AEdec(k,\ct)$ to denote the decryption of ciphertext $\ct$ under key $k$. Authenticated encryption can be instantiated with AES-GCM~\cite{AES:GCM}.

A digital signature scheme $\DS$ with key-generation algorithm $(\sk,\pk)\gets\DSgen(1^\sep)$, signing algorithm $\sigma\gets\DSsign(\sk, m)$, and verification algorithm $\{0,1\}\gets\DSverif(\pk,\sigma,m)$. In practice, $\DS$ can be instantiated with ECDSA~\cite{DBLP:journals/ijisec/JohnsonMV01}.

\subsection{Hypergeometric Distribution}
\label{sec:background:hypergeometric}

The hypergeometric distribution~$\hypergeom(N,C,n)$, where~$N$ is the overall number of elements in a set, $0\leq C \leq N$ is the number of ``special'' elements, and~$0\leq n \leq N$ is the number of extracted elements, without repetition, among all elements,
defines the probability of having a certain number of special elements in the extracted sample.
For a random variable~$X\sim \hypergeom(N,C,n)$, we have:
\begin{equation}
    \prob{X = x} = \frac{\binom{C}{x} \binom{N-C}{n-x}}{\binom{N}{n}}
    \quad\text{and}\quad
    \ExpVal{}{X} = n\frac{C}{N}.
\end{equation}
For $0 \leq t \leq n\frac{C}{N}$, the following tail bounds hold~\cite{skala2013hypergeometric}:
\begin{align}
    \label{hypergeometric:tail1}
    \prob{X \geq \ExpVal{}{X} + tn} &\leq e^{-2 t^2 n},\\
    \label{hypergeometric:tail2}
    \prob{X \leq \ExpVal{}{X} - tn} &\leq e^{-2 t^2 n}.
\end{align} 
\section{LiSA}
\label{sec:lisa}

\begin{table}[t!]
    \center
    \begin{tabular}{cl}
    \toprule
    Symbol & Description\\
    \midrule
    \multicolumn{2}{l}{\underline{\emph{System model}}}\\
    $\users$ & Set of users\\
    $\dusers$ & Set of dropped-out users\\
    $\cusers$ & Set of corrupted users\\
    $n$ & Overall number of users\\
    $\delta$ & Fraction of dropped-out users\\
    $\gamma$ & Fraction of corrupted users\\
    $Q$ & Public randomness\\
    \midrule
    \multicolumn{2}{l}{\underline{\emph{Protocol parameters}}}\\
    $\committee$ & Committee\\
    $\backupset_j$ & Backup neighbor set of committee member~$j$\\
    $k$ & Committee size, $k \leq n$\\
    $\ell$ & Size of backup-neighbor sets, $\ell \leq n$\\
    $t$ & Secret sharing reconstruction threshold, $t \leq \ell$ \\
    $\tilde{c}$ & Estimated number of corruptions in~$\committee$ \\
    $\alpha$ & Minimum fraction of aggregated inputs \\
    \bottomrule
    \end{tabular}
    \caption{Notation.}
    \label{table:parameters}
  \end{table}

\subsection{Problem description}

We assume a single server~$S$ and set of~$n$ users~$\users$. Without loss of generality, we identify each user with a member of $[n]$ so that $\users=[n]$. User $i\in\users$ holds private input $x_i$. The protocol should allow the server to learn $y=\sum_{i_\users} x_i$.

Users have no direct communication channels as all messages are routed through the server.
In line with previous work on single-server secure aggregation~\cite{DBLP:conf/ccs/BellBGL020,DBLP:conf/ccs/BonawitzIKMMPRS17}, we assume a Public Key Infrastructure (PKI) to distribute the genuine user public keys. Hence any pair of users can establish a confidential and authenticated channel. Further, each user has a confidential and authenticated channel with the server.

We seek to design protocols that are robust---that is, protocols that provide the aggregation output---in presence of a~$\delta$ fraction of users that can go offline at any time during the protocol execution.

\subsubsection{Threat Model}
We consider the exact same threat model of~\cite{DBLP:conf/ccs/BonawitzIKMMPRS17,DBLP:conf/ccs/BellBGL020}. In particular, we account for semi-honest as well as malicious settings. In the semi-honest settings, corrupted parties follow the protocol as expected; in the malicious settings, corrupted parties can act arbitrarily. In both cases, we allow the adversary to corrupt the server and a~$\gamma$ fraction of the users, statically. The security of our protocols guarantees that the input of an honest user is aggregated with at least an~$\alpha$ fraction of all other inputs before it is revealed to the server in the clear. In other words, the server learns $y=\sum_{i\in\users'}x_i$ only if
$\users'\subseteq\users$ and $|\users'|\geq \alpha n$.

Similar to previous work on SA~\cite{DBLP:conf/ccs/BonawitzIKMMPRS17,DBLP:conf/ccs/BellBGL020}, we consider denial of services attacks or attacks to the integrity of the computation as out of scope. Nevertheless, we show how to add integrity protection to \ours{} in Section~\ref{sec:discussion}. The notation used throughout the paper is summarized in Table~\ref{table:parameters}.

\subsection{Semi-honest Protocol}

We provide the specification of the semi-honest protocol in Figure~\ref{fig:protocol:both} (only text in black).

In our protocol, each user can assume up to three roles: regular user, committee member, and backup neighbor. We assume each user $i\in\users$ to have two key-pairs $(\sk_i,\pk_i)$ $(\ssk_i,\ppk_i)$ used for non-interactive key agreement. Key pair $(\ssk_i,\ppk_i)$ is used by user $i$ when she acts as a committee member, whereas key-pair $(\sk_i,\pk_i)$ is used for all other purposes.

During Round 1, all users and the server use the random beacon to select a \emph{committee} of users, denoted as $\committee$, at random. In particular, we use $\committee\gets\nchoosek(Q,\users,k)$ to denote a function that selects $k$ users from $\users$ by means of a PRG seeded with randomness~$Q$.

During Round 2, each user~$i$ adds noise to her input~$x_i$ by using $k$ random masks, each shared with one of the committee members. In particular, user~$i$ can agree on a shared key with committee member~$j$ by running $k^{*}_{i,j}=\KAprf(\textsc{"prg"};\sk_i,\ppk_j)$ and then obtain a mask as $F(k^{*}_{i,j})$, where $F$ is a PRG. The noisy input, denoted as $\ct_i$, is sent to the server.

The server aggregates the noisy inputs as $\ct_{agg}$, and asks each of the committee members for the aggregated randomness required to de-noise $c_{agg}$ (Round 3). Let $\users'_2$ be the set of users from who the server receives a noisy input. Each committee member $j$ runs a non-interactive key agreement with each user $i\in\users'_2$ to obtain $\{k^{*}_{j,i}=\KAprf(\ssk_j,\pk_i)\}_{i\in\users'_2}$. Next, committee member $j$ uses each of the agreed keys to obtain the shared masks by means of a PRF, aggregates all the shared masks as $\partial_j$ and sends it to the server. Finally, the server obtains the aggregated masks from each committee member and subtracts those to $\ct_{agg}$ to remove the noise and obtain the aggregated output.

The above protocol is not robust against dropouts by committee members. If a user $j\in\committee$ goes offline before she sends her aggregated noise share~$\partial_j$ to the server (Round 3, step 6), then the aggregated noisy input can no longer be de-noised. We add robustness by selecting a set~$\backupset_j$ of~$\ell$ \emph{backup neighbors} for each committee member $j$, and by asking $j$ to secret-share her secret key $\ssk_j$ among them, with reconstruction threshold $t\leq \ell$ (Round 2).

Similar to the committee selection, the selection of a backup neighborhood uses the randomness provided by the random beacon service. In this case, each committee member $j$ runs $\backupset_j\gets\nchoosek(Q||j,\users,l)$ to select $l$ backup neighbors.\footnote{We seed the PRG of the $\nchoosek$ function with $Q||j$---the random beacon $Q$ concatenated with the index of the committee member $j$---so that different committee members obtain independent sets of backup neighbors.}

In case a committee member~$j$ goes offline before she can send the aggregated noise share, the server can recover $j$'s secret key by asking for her shares to backup neighbors in $\backupset_j$ (Round 4). If $t$-many users in $\backupset_j$ send a share of $\ssk_j$ to the server (Round 5), then the server can recover the key and compute the aggregated noise share (Round 6), despite $j$ being offline.

Backup neighbors aid the server in recovering the secret key of a committee member only if the number of dropped-out committee members $|\committee_{drop}|$ is below $k-\tilde{c}$, where $\tilde{c}$ is the expected number of committee members that are corrupted. This is to make sure that even by corrupting some committee members and by recovering some secret keys via backup neighbors, the server is still missing at least one secret key of a committee member and, therefore, cannot de-noise individual noisy inputs of victim users.

Note that regular users need to be online only until Round 2, then they can go offline. Committee members and backup neighbors are required to remain online until Round 3 and until Round 5, respectively.

Also note that Figure~\ref{fig:protocol:both} assumes each user to encrypt a single message to ease readability. In case users encrypt vectors of size $m$, each element is encrypted separately, and aggregation is carried-out element-wise.
\begin{figure*}[th]
  \raggedright
  
  \textbf{Parties:} Server and users~$\users = [n]$.
  
  \textbf{Public parameters:} input domain~$\XX$, fraction of drop-outs~$\delta$, fraction of corruptions~$\gamma$, security parameter for cryptographic primitives $\sep$, committee size~$k$, backup-neighborhood size~$\ell$, secret sharing reconstruction threshold~$t$, minimum fraction of aggregated inputs~$\alpha$, maximum number of corrupt committee members~$\tilde{c}$.
  
  \textbf{Prerequisites:} Each user $i\in\users$ has key-pairs computed as $(\sk_i,\pk_i) \gets \KAgen(1^\sep)$, $(\ssk_i,\ppk_i) \gets \KAgen(1^\sep)$ \malicious{and $(\Ssk_i,\Spk_i)\gets\DSgen(1^\sep)$}; public keys $(\pk_i,\ppk_i\malicious{,\Spk_i})$ are registered with the PKI.
  
  For~$\users$ and~$r \in [6]$ and, we denote by~$\users_r$ the set of users that complete the execution of round~$r$ without dropping out, and we denote by~$\users_r '$ the set of users the server knows have completed round~$r$. It holds $\users'_r \subseteq \users_r \subseteq \users_{r-1}$ for all~$r\in [6]$.
  
  \smallskip
  \scalebox{0.8}{%
  \begin{minipage}{.6\textwidth}
  \textbf{Round 1}
  \begin{description}[resume]
      \item{Each party}
          \begin{enumerate}\setlength{\itemindent}{-.3in}
          \item receives random seed~$Q$
          \item selects committee $\committee\gets\nchoosek(Q,\users,k)$
          \end{enumerate}
  \end{description}
  \textbf{Round 2}
  \begin{description}[resume]
      \item{Committee member~$j \in \committee$}
          \begin{enumerate}\setlength{\itemindent}{-.3in}
          \item selects backup neighbors $\backupset_j\gets\nchoosek(Q||j,\users\setminus\{j\},\ell)$
          \item fetches public keys of backup neighbors $\{\pk_i\}_{i\in \backupset_j}$ from the PKI
          \item derives symmetric keys: $\{k^{e}_{j,i} \gets \KAenc(\textsc{"enc"};\sk_j,\pk_i)\}_{i\in \backupset_j}$\label{tag:hybrid1:1}
          \item secret shares key~$\ssk_j$: $\{S_{j,i}\}_{i\in\backupset_j}\gets\SSshare(\ell;t;\ssk_j)$
          \item encrypts shares of $\ssk_j$: $\{E_{j,i}\gets\AEenc(k^e_{j,i},S_{j,i})\}_{i\in\backupset_j}$\label{tag:hybrid3:1}
          \item sends $\{(j;i;E_{j,i})\}_{i\in\backupset_j}$  to the server
          \end{enumerate}
      \item{User~$i\in\users$}
          \begin{enumerate}\setlength{\itemindent}{-.3in}
          \item fetches public keys $\{\ppk_j\}_{j\in\committee}$ from the PKI
          \item derives symmetric keys:~$\{k^{*}_{i,j} \gets \KAprf(\textsc{"prg"};\sk_i,\ppk_j)\}_{j\in\committee}$\label{tag:hybrid7:1}
          \item computes blinded input: $\ct_i \gets x_i+\sum_{j\in\committee}F(k^{*}_{i,j})$
          \item \label{line:protocol:lazy:users} sends~$\ct_i$ to the server
          \end{enumerate}
  \end{description}
  
  \textbf{Round 3}
  \begin{description}
      \item{Server}
          \begin{enumerate}\setlength{\itemindent}{-.3in}
          \item receives encrypted key shares~$\{(j;i;E_{j,i})\}_{i\in \backupset_j}$  from $j\in\committee\cap\users'_2$ \\ and sends each of them to corresponding backup neighbor
          \item receive blinded inputs~$\ct_i$ from users~$i\in\users_2 '$
          \item aggregates input $c_{agg}\gets\sum_{i\in\users_2'}c_i$
          \item sends~$\users'_2$ to committee members $j\in \committee\cap\users'_2$.
          \end{enumerate}
      \item{Committee member $j\in\committee\cap\users'_2$}
          \begin{enumerate}\setlength{\itemindent}{-.3in}
            \item receives $\users'_2$ from the server
            \item if $\size{\users'_2}<\alpha n$, aborts
            \item fetches public keys $\{\pk_i\}_{i\in\users'_2}$ from the PKI
            \item derives symmetric keys~$\{k^{*}_{j,i} \gets \KAprf(\textsc{"prg"};\ssk_j,\pk_i)\}_{i\in\users'_2}$\label{tag:hybrid7:2}
            \item computes partial blinding $\partial_j\gets\sum_{i\in\users'_1}F(k^{*}_{j,i})$
            \item sends~$\partial_j$ to the server
          \end{enumerate}
          
  \end{description}

  \end{minipage}\quad\quad
  
  \begin{minipage}{.65\textwidth}
  
  \textbf{Round 4}
  
  \begin{description}
          \item{Server} \\
          \texttt{//let $\committee_{alive}:=\committee\cap \users'_3$ and $\committee_{drop} :=\committee \setminus \committee_{alive}$ and $\backupset = \bigcup_{j\in\committee} \backupset_j$}
          \begin{enumerate}\setlength{\itemindent}{-.3in}
          \item receives $\{\partial_j\}_{j\in\committee_{alive}}$  from committee members in~$\committee_{alive}$
          \item \label{line:committee:fully:alive} if $\size{\committee_{alive}} = k$ jumps to step~(\ref{line:M:final:decryption}) of Round 6
          \item sends~$\committee_{drop}$ to all users in $\backupset$\
          \end{enumerate}\setlength{\itemindent}{-.3in}
     \item{Backup neighbor~$i\in \backupset$}
          \begin{enumerate}\setlength{\itemindent}{-.3in}
          \item receives~$\committee_{drop}$ from the server
          \item \label{line:committee:too:may:dropouts} if $\size{\committee_{drop}} \geq k - \tilde{c}$ then aborts
          \item \malicious{computes signature $\sigma_i\gets\DSsign(\Ssk_i;\committee_{drop})$}
          \item \malicious{sends $\sigma_i$ to the server}
          \end{enumerate}
  \end{description}
  
  \textbf{Round 5}
  \begin{description}
      \item{Server}
          \begin{enumerate}\setlength{\itemindent}{-.3in}
              \item \malicious{receives $\{\sigma_i\}_{i\in \backupset \cap \users'_3}$ from users $\backupset \cap \users'_3$\\
                  and forwards them to all users in $\backupset \cap \users'_3$}
          \end{enumerate}
  
      \item{Backup neighbor~$i\in \backupset \cap \users'_3$}
          \begin{enumerate}\setlength{\itemindent}{-.3in}
          \item \malicious{receives $\{\sigma_i\}_{i\in \backupset \cap \users'_3}$}
          \item \malicious{fetches $\{\Spk_i\}_{i\in \backupset \cap \users'_3}$ from the PKI}
          \item \malicious{computes $\backupset_{ack} = \{ l\in\backupset\cap \users'_4 : \DSverif(\Spk_l;\sigma_l;\committee_{drop}) = \TRUE\}$}\label{tag:hybrid5:1}
          \item \malicious{if $\size{\backupset_j\cap\backupset_{ack}} < t$ for any~$j\in\committee$ then aborts}
          \item for any $j\in\committee_{drop}$ such that $i\in\backupset_j$,
              \begin{enumerate}\setlength{\itemindent}{-.3in}
              \item fetches $\pk_j$ from the PKI
              \item derives symmetric key~$(k^{e}_{i,j}) \gets \KAenc(\textsc{"enc"};\sk_i,\pk_j)$\label{tag:hybrid1:2}
              \item decrypts $S_{j,i} \gets \AEdec(k^e_{i,j};E_{j,i})$\label{tag:hybrid3:2}
              \item sends $S_{j,i}$ to the server\label{tag:hybrid4:2}
              \end{enumerate}
          \end{enumerate}
  \end{description}

  \textbf{Round 6}
  \begin{description}
      \item{Server}
          \begin{enumerate}\setlength{\itemindent}{-.3in}
              \item For each $j\in\committee_{drop}$
              \begin{enumerate}\setlength{\itemindent}{-.3in}
                  \item collects shares $\{S_{j,i}\}_{i\in\backupset_j\cap \users'_5}$ and aborts if receives less than $t$ shares
                  \item reconstructs secret key $\ssk_j\gets\SSrecon(\{S_{j,i}\}_{i\in\backupset_i\cap \users'_5})$
                  \item derives symmetric keys $\{k^{*}_{j,i}\gets \KAprf(\textsc{"prg"};\ssk_j;\pk_i)\}_{i\in\users_2'}$\label{tag:hybrid6:1}
                  \item computes missing partial blinding~$\partial'_j\gets\sum_{i\in\users'_2}F(k^{*}_{i,j})$
                  
              \end{enumerate}
                  \item \label{line:M:final:decryption} given $\{\partial_j\}_{j\in\committee}=\{\partial_j\}_{j\in\committee_{alive}} \cup \{\partial_j\}_{j\in\committee_{drop}}$, computes the output:\\
                      $y\gets c_{agg} - \sum_{j\in\committee}\partial_j$
                  
          \end{enumerate}

  \end{description}

  \end{minipage}%
  }
  
  \caption{{\ours} secure aggregation protocol. Instructions \malicious{highlighted in red} are executed only in the malicious version.}
  \label{fig:protocol:both}
  \end{figure*}

\subsection{Malicious Protocol}
The malicious protocol follows the same blueprint of the semi-honest one. Differences between the two protocol stems from the ability of a corrupted server to act arbitrarily. The text in red in Figure~\ref{fig:protocol:both} shows the changes we make to the semi-honest protocol so to withstand a malicious server.

A malicious server may declare different sets of dropped-out committee members, each of size $<k-\tilde{c}$, to different backup neighborhoods, so to obtain their key-shares. If the server manages to obtain the secret keys of each honest committee member, it can remove noise from any single noisy input, thereby breaking security. Hence, we use Round 4 and Round 5 to run a consistency check among backup neighbors, so to agree on the set of dropped-out committee members; if the size of this set  is bigger than threshold $k-\tilde{c}$, then no honest backup neighbor will help the server in recovering the missing keys.

\section{Analysis}
\label{sec:analysis}

In this section we prove that {\ours} is correct, i.e., the server obtains the aggregation of users provided inputs, as long as all parties follow the protocol and at most~$\delta n$ users dropout. We also prove that {\ours} is secure, i.e., if the protocol execution completes then the server learns the aggregation of no less than~$\alpha n$ inputs, despite a malicious server and up to~$\gamma n$ compromised users.

\paragraph{Overview}
We will analyze the probability of certain events occurring during an execution of the secure aggregation protocol, and relate them to the probability of violating correctness or security.

Let $\cusers$ be the set of corrupted users and let $\dusers$ be the set of users that drop out. Further, let~$d$, $d_j$, $c$, and~$c_j$ denote the number of dropped committee members, the number of dropped backup neighbors of committee member~$j$, the number of corrupt committee members, and the number of corrupt backup neighbors of committee member~$j$, respectively.  That is:
\begin{equation*}
d := \size{\committee \cap \dusers},
\quad
d_j := \size{\backupset_j \cap \dusers},
\quad
c := \size{\committee \cap \cusers},
\quad
c_j := \size{\backupset_j \cap \cusers}.
\end{equation*}

Recall that {\ours} (pseudo-)randomly selects committee members and backup neighbors, based on a public random seed. As users in the committee, as well as in each backup neighborhood, must be distinct, we can think of selecting them, without replacement, from a population of~$n$ users.
Then, in the worst case with all~$\delta n$ users dropping out and all~$\gamma n$ users being corrupt, i.e., $\size{\dusers} = \delta n$ and $\size{\cusers} = \gamma n$, random variables~$d$, $d_j$, $c$ and~$c_j$ are distributed according the hypergeometric distribution (cf.~Section~\ref{sec:background:hypergeometric}):
\begin{gather}
    d \sim \hypergeom(n,\delta n, k),\quad d_j \sim \hypergeom(n-1,\delta n, \ell),\label{eq:d:hyper}\\
    c \sim \hypergeom(n,\gamma n, k),\quad c_j \sim \hypergeom(n-1,\gamma n, \ell).
\end{gather}

\subsection{Correctness}
\label{sec:analysis:correctness}

In addition to the correctness of the cryptographic building blocks of~{\ours}, we need to define some requirements to guarantee robustness.
This means limiting the number of drop-outs in the committee and in the backup neighborhoods, as specified through the following events.

\smallskip

    \noindent\textit{Event $C_1$}: all committee members stay alive
    \begin{equation}
        \size{\committee\setminus \dusers} = k.
    \end{equation}

    \noindent\textit{Event $C_2$}: sufficiently many committee members stay alive:
    \begin{equation}
        \size{\committee\setminus\dusers} > \tilde{c}.
    \end{equation}

    \noindent\textit{Event $C_3$}: sufficiently many backup neighbors of dropped committee members stay alive:
    \begin{equation}
        \forall j\in \committee\cap\dusers\ :\ \size{\backupset_j \setminus \dusers} \geq t.
    \end{equation}

\smallskip

Correctness of {\ours} requires that either all committee members complete the protocol (event $C_1$),
or for each committee member~$j$ that dropped out, the server can recover enough shares from neighbors in $\backupset_j$ to compute the secret key(s) of $j$ (events $C_2$ and $C_3$).
Therefore, it is sufficient to select parameters~$k$, $\ell$, and~$t$ (based on~$n$ and $\delta$) so that the event $C_1 \lor (C_2 \land C_3)$ occurs with overwhelming probability.
We can ensure these requirements are met with high probability by selecting parameters~$k$, $\ell$, and~$t$ such that
\begin{equation}
    \prob{C_1 \lor (C_2 \land C_3)} \geq 1 - 2^{-\eta}.
\end{equation}
To this end, we consider the complement event:
\begin{equation}
    \prob{C_1 \lor (C_2 \land C_3)} \geq 1 - 2^{-\eta} \Longleftrightarrow
    \prob{\neg C_1 \land \neg(C_2 \land C_3)}  \leq 2^{-\eta}
\end{equation}
where
\begin{align}
    \prob{\neg C_1 \land \neg(C_2 \land C_3)}
    = & \prob{\neg C_1 \land (\neg C_2 \lor \neg C_3)}\\
    = & \prob{ (\neg C_1 \land \neg C_2)\lor (\neg C_1 \land \neg C_3)}\\
    \leq & \prob{ \neg C_1 \land \neg C_2} + \prob{\neg C_1 \land \neg C_3}.
\end{align}
By observing that
\begin{align}
    \neg C_1 &:  \size{\committee\setminus \dusers} < k \Longleftrightarrow \size{\committee\cap\dusers}> 0 \\
    \neg C_2 &: \size{\committee\setminus\dusers} \leq \tilde{c} \Longleftrightarrow \size{\committee\cap\dusers} \geq k - \tilde{c}\\
    \neg C_3 &: \exists j\in \committee\cap\dusers\ :\ \size{\backupset_j \setminus \dusers} < t\\
    & \Longleftrightarrow \exists j\in \committee\cap\dusers\ :\ \size{\backupset_j \cap \dusers} \geq \ell - t
\end{align}
we obtain
\begin{align}
    \prob{ \neg C_1 \land \neg C_2}
    & = \prob{ (d > 0) \land (d \geq k - \tilde{c})}\\
    & = \prob{d \geq k - \tilde{c}}
\end{align}
and
\begin{align}
    \prob{\neg C_2 \land \neg C_3}
    &= \prob{(d > 0) \land \exists j\in \size{\committee\cap\dusers}: d_j \geq\ell -t}\\
    &\leq \prob{ \exists j\in \committee: d_j \geq \ell -t}\\
    &\leq \sum_{j\in\committee} \prob{d_j \geq \ell -t}
\end{align}
Putting it all together, we require:
\begin{equation}
    \prob{ d \geq k - \tilde{c}} + \sum_{j\in\committee} \prob{d_j \geq \ell -t} \leq 2^{-\eta}.
\end{equation}
Let $X\sim \hypergeom(n,\delta n, k)$ and~$Y \sim \hypergeom(n-1,\delta n, \ell)$.
By Equation~\eqref{eq:d:hyper}, we can express the requirement above as:
\begin{equation}
\label{eq:correctness:condition:hypergeometric}
    \prob{ X \geq k - \tilde{c}} + k\ \prob{Y \geq \ell -t} \leq 2^{-\eta}.
\end{equation}
The following two conditions imply equation~\eqref{eq:correctness:condition:hypergeometric}:
\begin{align}
    \prob{ X \geq k - \tilde{c}}   &\leq 2^{-(\eta +1)}\\
    k\ \prob{Y \geq \ell -t}       &\leq 2^{-(\eta+1)}
\end{align}
Let~$\tilde{c} = \tilde{\gamma} k$, with $\gamma < \tilde{\gamma} < 1-\delta$.
Using the tail bounds of the hypergeometric distribution (cf.~Section~\ref{sec:background:hypergeometric}), we have
\begin{align}
    \prob{ X \geq k - \tilde{c}} &= \prob{ X > \delta k + (1-\delta) k - \tilde{\gamma} k} \\
                                 &= \prob{ X > \ExpVal{}{X} + (1-\delta-\tilde{\gamma}) k } \\
                                 &\leq e^{-2(1-\delta -\tilde{\gamma})^2k},
\end{align}
and similarly
\begin{align}
    \prob{Y \geq \ell -t}   &= \prob{Y \geq \frac{\delta n}{n-1}\ell +\left(1-\frac{\delta n}{n-1}\right)\ell -\frac{t}{\ell}\ell}\\
                            &= \prob{Y \geq \ExpVal{}{Y} +\left(1-\frac{\delta n}{n-1}-\beta\right)\ell}\\
                            &\leq e^{-2(1-\frac{\delta n}{n-1} - \beta)^2\ell}.
\end{align}
with $t = \beta \ell$ and $\beta < 1 - \delta$.
We therefore require
\begin{equation}
    e^{-2(1-\delta-\tilde{\gamma})^2k} <  2^{-(\eta +1)} \Longrightarrow k > \frac{1}{2\log(e)} \frac{\eta +1}{(1 - \delta - \tilde{\gamma})^2}
\end{equation}
and
\begin{equation}
    k\ e^{-2(1-\delta - \frac{t}{\ell})^2\ell} < 2^{-(\eta+1)} \Longrightarrow \ell >\frac{\log(k)+\eta +1}{2 \log(e)(1-\frac{\delta n}{n-1}-\beta)^2}
\end{equation}

We finally obtain
\begin{align}
    k    & > \frac{1}{2\log(e)} \frac{\eta +1}{(1 - \delta - \tilde{\gamma})^2}\\
    \ell &>\frac{\log(k)+\eta +1}{2 \log(e)(1-\frac{\delta n}{n-1}-\beta)^2}
\end{align}

All cryptographic primitives invoked by {\ours} are known to be correct. Therefore, we can state the following result.

\begin{theorem}[Correctness]
    Consider an execution of {\ours} with user inputs~$\{x_i\}_{i\in\users}$.
    If no more than~$\delta n$ users dropout throughout the execution, i.e., $\size{\users'_5} \geq (1-\delta)n$,
    then the protocol completes and the server obtains output~$y = \sum_{i\in\users'_1} x_i$ with overwhelming probability.
\end{theorem}

\subsection{Security}
\label{sec:analysis:security}

Similarly to the case of correctness, we identify relevant events that establish a limit on the number of corrupt users in the committee and in the backup neighborhoods.

Intuitively, if the protocol completes at least one committee member must be honest and alive.
Otherwise, the server would obtain the decryption keys of the~$c$ committee members it corrupts, and it would also obtain the decryption keys of the~$k-c$ dropped-out committee members from the corrupted backup neighbors. Once the server has the secret keys of all committee members, it can decrypt any individual ciphertext, thereby break security. Hence, we must ensure the following event:
\begin{equation}
    \label{eq:security:requirement:1}
    \size{\committee \cap \cusers} + \size{\committee\cap \dusers} < \size{\committee}.
\end{equation}

To meet this requirement, our protocol instructs backup neighbors to release shares of committee members' decryption keys only if not too many of them have dropped out, where ``too many'' is set as
\begin{equation}
    \size{\committee\cap \dusers} < k - \tilde{c}.
\end{equation}
Combining this with the requirement in equation~\eqref{eq:security:requirement:1}, we have:
\begin{equation}
    \size{\committee \cap \cusers} < \size{\committee} - \size{\committee\cap \dusers} = k - (k - \tilde{c}) = \tilde{c}
\end{equation}

Moreover, the protocol must ensure that backup neighbors do not release (too many) shares of honest and alive committee members.
The reason is that the malicious server could equivocate during the consistency check, and try to obtain the decryption keys of more than~$k - \tilde{c} - 1$ committee members by presenting different sets of dropped committee members to different backup neighbors.

For a victim $j\in \committee$, consider two subsets of her backup neighbors~$S_1,S_2\subseteq \backupset_j$ of size $\size{S_1} = \size{S_2} = t$ and that overlap in corrupt members: $S_1 \cap S_2 = \cusers\cap \backupset_j$.
The server could present to~$S_1$ a set of dropped users~$\committee_{drop,1}$ such that~$j\in\committee_{drop,1}$, and to~$S_2$ a set~$\committee_{drop,2}$ such that~$j\notin\committee_{drop,2}$.
In this way, the server would obtain~$t$ shares~$s_{j,i}$ from the backup neighbors~$i\in S_1$ and reconstruct key~$\sk_j$, and would also obtain~$t$ signatures acknowledging that~$j$ is alive from the backup neighbors in~$S_2$,  which would allow the server to reconstruct more decryption keys than we want (i.e., more than $k - \tilde{c} -1$).
To prevent this, we require that any two such sets~$S_1$ and~$S_2$ overlap in at least an honest member:
\begin{equation}
    \ell < \size{S_1\cup S_2},
\end{equation}
and since we chose $S_1$ and $S_2$ to be of size~$t$ each and with at least~$c_j$ overlapping members, we have
\begin{equation}
    \size{S_1\cup S_2} \leq 2t - c_j.
\end{equation}
Combining the results, we finally obtain
\begin{equation}
    2t - c_j > \ell \Longleftrightarrow c_j < 2t - \ell.
\end{equation}

We therefore establish the following security requirements.

    \noindent\textit{Event $S_1$:} Not too many corrupt committee members:
    \begin{equation}
        \size{\committee\cap \cusers} \leq \tilde{c}.
    \end{equation}
    \noindent\textit{Event $S_2$:} Not too many corrupt backup neighbors:
    \begin{equation}
        \forall j\in \committee: \size{\backupset_j\cap\cusers} \leq 2t -\ell.
    \end{equation}

Security of {\ours} requires $S1 \land S2$.
By selecting parameters~$k$, $\ell$, and~$t$ (based on~$n$ and $\gamma$) we can ensure that this event occurs with overwhelming probability.

Again considering the complementary event, we have:
\begin{equation}
    \prob{S_1 \land S_2} \geq 1 - 2^{-\sep} \Longleftrightarrow
    \prob{\neg S_1 \lor \neg S_2}  \leq 2^{-\sep}
\end{equation}
We thus have
\begin{align*}
    \prob{\neg S_1 \lor \neg S_2}
    &\leq \prob{\neg S_1}+\prob{\neg S_2}\\
    & \leq \prob{c > \tilde c} + \prob{\cup_{j\in\committee}c_j > 2t - \ell}\\
    & \prob{c > \tilde c} + \sum_{j\in\committee}\prob{c_j > 2t - \ell}
\end{align*}
Now, let $Z\sim\hypergeom(n,\gamma n, k)$ and $W\sim\hypergeom(n-1,\gamma n,\ell)$.
Then~$c \sim Z$ and $c_j \sim W$ for all~$j\in\committee$, and we can express the above requirement as
\begin{equation}
    \label{eq:security:condition:hypergeometric}
    \prob{Z > \tilde c} + k\ \prob{W > 2t - \ell} \leq 2^{-\sep}
\end{equation}
The following two conditions imply equation~\eqref{eq:security:condition:hypergeometric}:
\begin{align}
    \prob{ Z \geq \tilde{c}}   &\leq 2^{-(\sep +1)}\\
    k\ \prob{W \geq 2t-\ell}       &\leq 2^{-(\sep+1)}
\end{align}
Again using the tail bounds of the hypergeometric distribution, and setting $\tilde{c} = \tilde{\gamma}k$, we find:
\begin{align}
    \prob{ Z \geq \tilde{c}}
    &= \prob{Z \geq \gamma k - \gamma k + \tilde{\gamma}k }\\
    &= \prob{Z \geq \ExpVal{}{Z} + (\tilde{\gamma} - \gamma)k }\\
    &\leq e^{-2(\tilde{\gamma}-\gamma)^2k}
\end{align}
and by requiring $e^{-2(\tilde{\gamma}-\gamma)^2k} < 2^{-(\sep +1)}$ we finally obtain
\begin{equation}
    k > \frac{\sep + 1}{2 \log(e)(\tilde{\gamma} - \gamma)^2}.
\end{equation}
Similarly, we require:
\begin{align}
    \prob{W \geq 2t-\ell}
    &= \prob{W \geq \frac{\gamma n}{n-1} \ell - \frac{\gamma n}{n-1} \ell + 2 (\beta -1) \ell}\\
    &= \prob{W \geq \ExpVal{}{W} - (2\beta -\frac{\gamma n}{n-1} -1)\ell}\\
    &\leq e^{-2(2\beta -\frac{\gamma n}{n-1} -1)^2 \ell}
\end{align}
with $t = \beta \ell$ and $\frac{1}{2}(1+\frac{\gamma n}{n-1})<\beta$.
By requiring $k\ e^{-2(2\beta -\frac{\gamma n}{n-1} -1)^2 \ell} < 2^{-(\sep + 1)}$ we finally obtain
\begin{equation}
    \ell > \frac{\log(k)+ \sep + 1}{2\log(e)(2\beta -\frac{\gamma n}{n-1} -1)^2}
\end{equation}

As $\frac{1}{2}(1+\frac{\gamma n}{n-1})<\beta<1-\delta$, we require that~$1-2\delta-\frac{\gamma n}{n-1}>0$. 

Putting it all together, we can select parameters~$k$, $\ell$ and~$t$ so that~{\ours} guarantees the security requirements are fulfilled.

To formally prove that our protocol (instantiated with~$k$, $\ell$ and~$t$ as above) provides privacy of users' inputs against a malicious adversary controlling the server and a subset~$\cusers$ of users, we show how to construct a simulator whose output is computationally indistinguishable from the output of any such malicious adversary.
The simulator is given as input only the protocol leakage, e.g., the output of the server, and simulates a faithful protocol execution for the adversary.
We provide a simulation-based security proof in Appendix~\ref{sec:simulation:proofs}.

\section{Performance Evaluation}
\label{sec:performance}

\subsection{Theoretical evaluation}

We theoretically evaluate the communication of users and the server for the malicious version of our protocol.

\paragraph{Regular users: $\asim (m+k)$ communication.} Regular users --- those not selected as committee members nor as backup neighbors--- must be online only during the first two rounds. They fetch public keys of $k$ committee members from the PKI and send $m$ noisy inputs.

\paragraph{Committee members: $\asim (n+m +\ell)$ communication.} Each committee member fetches $\ell$ public keys from the PKI and sends $\ell$ ciphertexts Round 2. Next, in Round 3, receives up to $n$ public keys and sends $m$ random values.

\paragraph{Backup neighbor: $\asim (k\ell)$ communication.} A user can be selected as a backup neighbor by multiple committee members. Hence, it receives up to $k$ ciphertexts during Round 3. In Round 4, each backup neighbor receives one messages from the server and sends one signature. Next, during Round 5 it receives up to $k\ell$ signatures and sends up to $k$ shares.

\paragraph{Server: $\asim (nm+k^2\ell^2)$ communication.} The server receives $k\ell$ shares and $nm$ blinded inputs during Round 3, and sends $k\ell$ shares. During Round 4, the server receives up to $km$ partial blindings and sends one message to up to $k\ell$ users. During Round 5, it receives up to $k\ell$ signatures and sends the same number of signatures to each member of each backup neighborhood. Finally, during round 6 it receives up to $k\ell$ shares.

The server aggregates up to $m$ partial blindings. In case committee members drop out, the server makes up to $k$ calls to $\SSrecon$, up to $nm$ calls to $\KA$, and up to $nm$ partial blindings recoveries.

\smallskip

Table~\ref{table:theoretic} summarizes asymptotic communication overhead for each type of party. For completeness, the table also include the overhead for the semi-honest protocol and the number of rounds. As demonstrated in Section~\ref{sec:analysis}, $k$ and $\ell$ are $\asim (1)$, it is sufficient that:
\begin{align*}
    k    & > \max{\left(\frac{1}{2\log(e)} \frac{\eta +1}{(1 - \delta - \tilde{\gamma})^2},\frac{1}{2\log(e)} \frac{\sep +1}{(\tilde{\gamma}-\gamma)^2}\right)}\\
    \ell &> \max{\left(\frac{\log(k)+\eta +1}{2 \log(e)(1-\frac{\delta n}{n-1}-\beta)^2},\frac{\log(k)+\sep +1}{2 \log(e)\left(2\beta-1-\frac{\gamma n}{n-1}\right)^2}\right)}
\end{align*}

The communication overhead for a user is consequently reduced from $\asim (k+m)$ to $\asim (m)$. Communication overheads of committee members, of backup neighbors and of the server are similarly formulated.

\begin{table}
\begin{center}
\begin{tabular}{lllc}
  \hline
  &\textbf{Semi-honest}&\textbf{Malicious}&\textbf{\# Rounds}\\
  \hline
  \hline
  User      & $\asim(m)$      & $\asim(m)$   & 2   \\
  Committee & $\asim(n+m)$      & $\asim(n+m)$  & 3\\
  Backup    & $\asim(1)$        & $\asim(1)$     & 5  \\
  Server    & $\asim(nm)$    & $\asim(nm)$ & 6\\
  \hline
\end{tabular}
\caption{Theoretical communication overhead.}
\label{table:theoretic}
\end{center}
\end{table}

\subsection{Numerical parameter optimization}

{\ours} takes only $2$ rounds for regular users and~$3$ rounds for committee members. Backup neighbors are required to be online for $5$ out of $6$ rounds, but only in case any committee member drops out before the third round is over. In the following section, we show how many committee members~($k$) and how many backup neighbors~($\ell$) are needed to satisfy the security and correctness requirements, given~$n$ users, corruption rate~$\gamma$, and dropout rate~$\delta$. We will show that the vast majority of users act as regular users and thus can complete the protocol in~$2$ rounds with little overhead.
We also compare our committee and neighborhood sizes with the neighborhood size of Bell~\etal~\cite{DBLP:conf/ccs/BellBGL020} (``Bell'' from now on).
To this end, we wrote a program in Julia \cite{DBLP:journals/siamrev/BezansonEKS17} to obtain the minimum committee size $k$ and number of backup neighbors $\ell$ for {\ours} to satisfy security and correctness conditions, both in the semi-honest and malicious cases. Reported figures for Bell are estimated from the results presented in~\cite{DBLP:conf/ccs/BellBGL020}.

\begin{figure}[t]
    \includegraphics[width=1.0\linewidth,left]{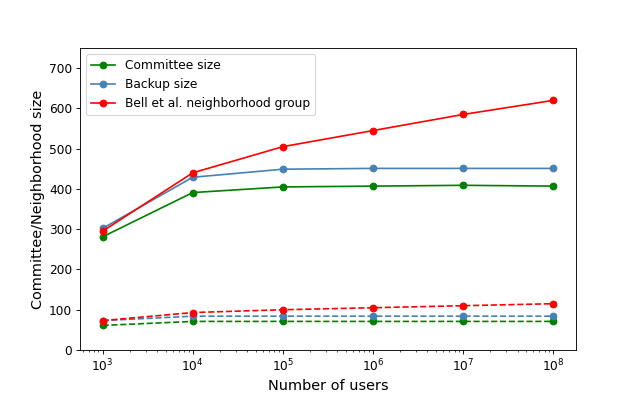}
    \caption{Semi-honest settings. Size of committee and size of backup neighborhood for {\ours}, and size of neighborhood for Bell~\etal. Solid lines assume $\gamma=0.33$ and $\delta=0.33$; dashed lines assume $\gamma=0.05$ and $\delta=0.33$. We set $\sigma=40$ and $\eta=20$.}
    \label{fig:hbc0305}
\end{figure}

Figure~\ref{fig:hbc0305} shows committee size~$k$ and backup neighborhood size~$\ell$ in the semi-honest settings. In the same figure, we also report the neighborhood size for Bell.
As $n$ grows, both~$k$ and~$\ell$ parameters for {\ours} remain almost constant; the neighborhood size for Bell grows but the slope is less steep for large number of users ($10^5$ and above).
For $n=10^6$ users, corruption rate $\gamma=0.33$ and dropout rate $\delta=0.33$, {\ours} requires only $k=407$ committee members, each having $\ell=451$ backup neighbors. In other words, more than $81\%$ of the users are 
regular users and must be online only for $2$ rounds. When $\gamma=0.05$ and $\delta=0.33$, this number increases to 99\%. Using the same parameters, Bell requires each user to have a backup neighborhood of more than 550 and 100 users respectively. All users are required to be online for the whole protocol execution ($5.5$ rounds).

\begin{figure}[t]
    \includegraphics[width=1.0\linewidth,left]{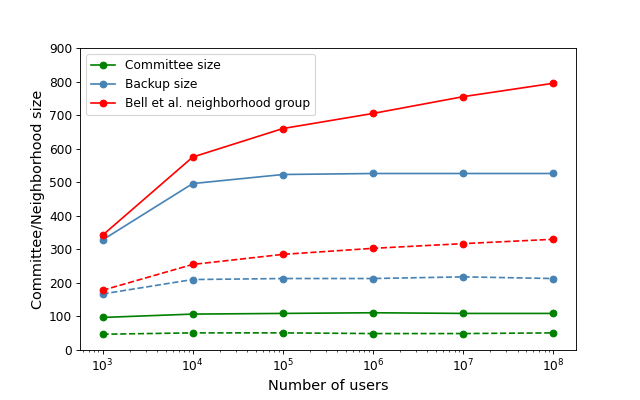}
    \caption{Malicious settings. Size of committee and size of backup neighborhood for {\ours}, and size of neighborhood for Bell~\etal. Solid lines assume $\gamma=0.2$ and $\delta=0.2$; dashed lines assume $\gamma=0.05$ and $\delta=0.2$. We set $\sigma=40$ and $\eta=30$.}
    \label{fig:0205}
\end{figure}

Figure~\ref{fig:0205} shows the sizes of relevant sets in the malicious settings. In this scenario, the backup neighborhood size~$\ell$ increases at first, until reaching a constant value, and is always smaller than the size of a neighborhood in Bell. For $n=10^6$ and $\gamma=\delta=0.2$, {\ours} uses~111 committee members, each with 526 backup neighbors. Only $0.01\%$ of the users are committee members, and less than $5.4\%$ are backup neighbors.

In our protocol, the security and correctness requirements for the committee are the same in both the honest-but-curious and the malicious model, i.e., the committee size stays the same. The security requirements for the backup neighbors are relaxed in the honest-but-curious protocol, allowing a smaller number of backup neighbors per committee member than in the malicious case.

\subsection{Benchmarks}

\begin{figure*}[th!]
\centering
    \begin{subfigure}[t]{0.33\textwidth}
    \centering
    \includegraphics[width=1.1\linewidth,left]{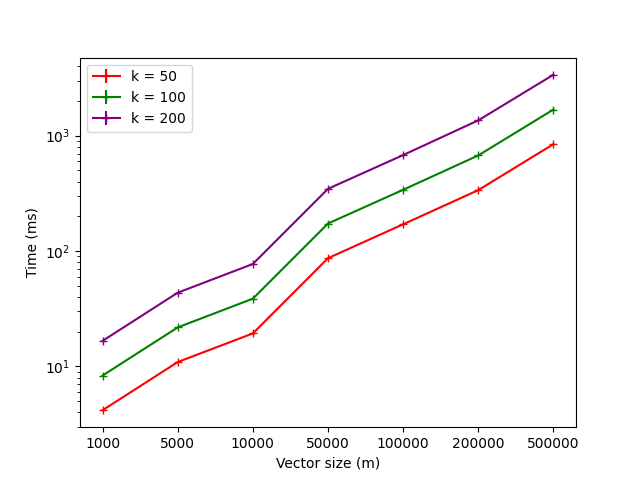}
    \caption{Regular user}
    \label{fig:perf:client}
   \end{subfigure}
    \begin{subfigure}[t]{0.33\textwidth}
    \centering
    \includegraphics[width=1.1\linewidth,left]{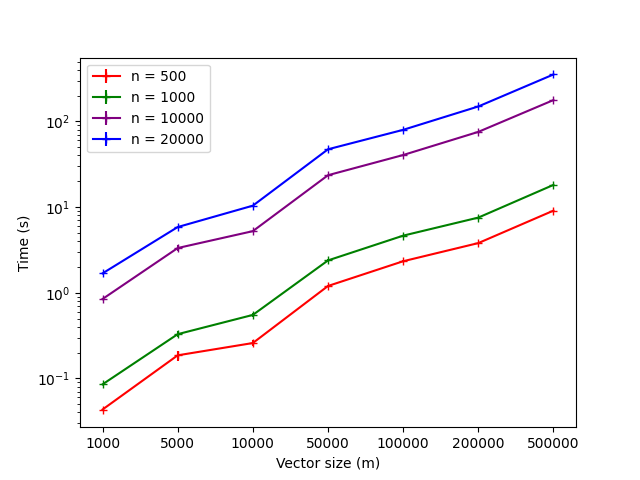}
    \caption{Committee member}
    \label{fig:perf:committee}
    \end{subfigure}
    \begin{subfigure}[t]{0.33\textwidth}
    \centering
    \includegraphics[width=1.1\linewidth,left]{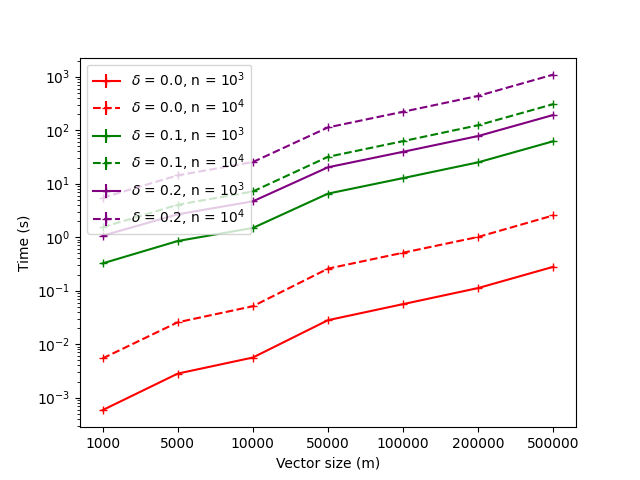}
    \caption{Server}
    \label{fig:perf:server}
    \end{subfigure}
 \caption{Performance of {\ours} in terms of average runtime per participant, as the input size    increases. We consider as participant's roles: regular users, for different committee sizes~$k$ (Fig.~\ref{fig:perf:client}); committee member, for different numbers of users~$n$ (Fig~\ref{fig:perf:committee}); and server, for different numbers of users~$n$ and dropout rates~$\delta$.
    Solid and dashed lines represent~1k users and~10k users respectively.
    }
 \label{fig:performance}
 \end{figure*}

To assess the performance of {\ours}, we implemented and tested a prototype written in Golang 1.20 that utilizes the necessary cryptographic libraries, including crypto/aes, crypto/ecdsa, crypto/elliptic, crypto/sha256 and golang.org/x/crypto/hkdf, to ensure secure and efficient cryptographic operations.
All experiments were run on a machine with Intel(R) Core(TM) i5-9600K CPU @ 3.70GHz and 62~GB of RAM, running Ubuntu~18.04 operating system. Our results are summarized in Figure~\ref{fig:performance}.

Figure~\ref{fig:perf:client} shows the time for a user to compute a blinded input (Round 2) as the size of the input vector increases, for different committee sizes. For example, the overhead for a vector of size~$m=100K$ ranges between~170ms ($k=50$) and 340ms ($k=100$). Note that user overhead is only dependent on~$m$ and $k$. The user overhead in Bonawitz~\etal~\cite{DBLP:conf/ccs/BonawitzIKMMPRS17} is dependent on~$m$ and~$n$. For example, Bonawitz~\etal~report $694ms$ to compute the noisy input for a vector of~$100K$ elements with $n=500$, and $1357ms$ for $n=1000$.

Figure~\ref{fig:perf:committee} shows the time for a committee member to compute the aggregated randomness in Round 3. This time is only dependent on the number of elements in the input vector $m$ and the number of users $n$.
For example, The overhead for a vector of size $m=100K$ ranges between 1s ($n=1K$) and 42s ($n=10K$).

The server runtime for the entire aggregation process is shown in Figure~\ref{fig:perf:server}. The server runtime depends on the dropout rate~$\delta$, as in Round 6 the server needs to reconstruct the missing partial binding for the offline committee members.
The computation cost of the server increases as the number of users grows. For example, for an input with~$m=100K$ entries and $n=10K$ users, the server requires approximately~80s and~210s for~10\% and~20\% dropouts, respectively. 
For the same input size ($m=100K$) but with only $n=500$ users and a dropout rate of~10\%, the server overhead in~\cite{DBLP:conf/ccs/BonawitzIKMMPRS17} is approximately 62 seconds. We note that~\cite{DBLP:conf/ccs/BonawitzIKMMPRS17} does not consider larger values of $n$.

\subsection{Integration with FL applications}
\label{sec:integration:FL:application}

To demonstrate the usefulness of our proposal for realistic federated-learning applications,
we integrated {\ours} into a concrete FL training algorithm and evaluated its effectiveness.
We compare the performance of {\ours} with the default (non-private) aggregation. Recall that standard aggregation generates the aggregated model-update by averaging the local updates provided by users, while secure aggregation enhances it by providing cryptographic protection to the user updates (i.e., {\ours} adds a pseudorandom mask to each input component).

\paragraph{Dataset.}
We use the Bank Marketing Data Set~\cite{DBLP:journals/dss/MoroCR14}, a dataset containing real data related to direct marketing campaigns based on phone calls to users.
The dataset contains 41,188 instances of bank user data, including 20 features such as age, job, marital status, education, and previous marketing campaigns, and the contact outcome (``success'' or ``failure'') as the label. We normalized the dataset to ensure better performance and accuracy of the model.
The dataset is unbalanced, as only 4,640 out of the 41,188 are successful.
We divided the total dataset into two, 80\% training data and 20\% testing data, with 32,950 records and 8,238 records respectively.

\begin{table}[t!]
    \begin{center}
    \begin{tabular}{ccccc}
        \hline
        &\textbf{\# users} & \textbf{\# rounds}  & \textbf{Runtime (s)} &\textbf{Accuracy} \\
        \hline
        \hline
        Baseline & \multirow{2}{*}{100} & 7 & 72.11 & 98.45\% \\
        \ours & & 7 & 74.66 & 98.23\% \\
        \hline
        Baseline  &\multirow{2}{*}{500} & 8 &  92.44 & 98.23\% \\
        \ours &  & 9 & 133.76 &  97.13\% \\
        \hline
        Baseline  & \multirow{2}{*}{1000} & 7 & 59.79 & 97.12\% \\
        \ours & & 11 &  158.05 &  97.34\% \\
        \hline
        \end{tabular}
    \end{center}
    \caption{Performance comparison of standard aggregation  vs.~{\ours} secure aggregation in a federated learning application. We report the number of rounds until convergence, overall training time, and accuracy.}
    \label{table:integration performance}
    \end{table}

\paragraph{Setup.}
We implemented a federated learning process using Golang 1.20, based on an open-source machine learning library~\cite{MLP:implementation}.
Each training round consists of a local training phase run by each user, with a fixed learning rate of~0.01 and a pre-specified number of local training epochs set to 10, and an aggregation phase to combine the local models.
Users train their local model using a multi-layer perceptron with three layers: an input layer with 62 neurons, a hidden layer with 40 neurons, and an output layer with 2 neurons~\cite{MLP:data_clearning}. We used the normal sigmoid function as the activation function.
To emulate a realistic FL deployment, we horizontally split the training set and conducted multiple rounds of training and aggregation until convergence, i.e., until the model stabilizes.
We ran various executions of the aforementioned FL process, varying the number of users $n\in\{100, 500, 1000\}$.
We repeated each experiment~5 times.

With this set of experiments, we aim to demonstrate the feasibility of embedding~{\ours} into FL deployments, hence we set the dropout and corruption rates to $\delta = \gamma = 0$ for the sake of simplicity.
To evaluate the effectiveness of our proposal,  we compare the cases where the aggregation phase uses (i) standard aggregation (i.e., without privacy) vs.~(ii)  {\ours}, and we measure in each case: minimum number of rounds needed to converge, overall training time, and prediction accuracy of the resulting model. The results of our experiments are summarized in Table~\ref{table:integration performance}.

\paragraph{Results.}
Our results show that using our secure aggregation protocol slightly increases the minimum number of rounds required for the model to converge when the number of users exceed~$100$ (precisely, from 8 to 9 rounds for $n = 500$, and from 7 to 11 rounds for $n = 1000$), compared to the default aggregation protocol. This behavior can be attributed to precision loss introduced by {\ours}.
Besides, the total training time is higher for {\ours} compared to the baseline (requiring in addition~$\approx 2s$, $41s$, and $98s$ respectively for $n = 100$, $500$, and $1000$ respectively), due to the extra blinding and unblinding steps in the aggregation phase.
Despite this, our protocol has a negligible impact on the accuracy and F1 score of the model, demonstrating that {\ours} can improve the privacy provisions of federated learning without sacrificing accuracy.

\subsection{Simulated Results}
In case of federated learning applications that use secure aggregation, the application runs across several rounds and invokes the aggregation protocol at every round. If all users have the same per-round overhead, it is easy to compute the overhead across several rounds. In {\ours}, however, a user may have different overhead, depending on the role it takes in a round (e.g., regular user, committee member, backup neighborhood). Thus, it is fair to consider user overhead across multiple rounds.

To this end, we wrote a python program that simulates several rounds of {\ours} and collects statistics for each of the users.
In this set of experiments, we set $n\in\{10^3,10^4,10^5,10^6\}$, $\gamma=\delta=0.2$, and run the secure aggregation protocol for~100 consecutive rounds. Table~\ref{table:simComBack} shows the measured average number of times a single user was selected as a committee member or as a backup neighbor during the 100 executions of the protocol.

For example, with $n=10^5$, the maximum number of times one user was selected as a committee member was $3$, and $90.4\%$ of users where never selected as part of the committee. The average number of times a user was selected as a backup neighbor was $52 \pm 7$. The maximum number of times a user was selected as a backup neighbor was $88$ (over 100 rounds).
Note however that a user can be selected as a backup neighbor multiple times, one by each committee member, in a given round.

\begin{table}[h]
\begin{center}
\begin{tabular}{lrr}
  \hline
   &\textbf{Committee}&\textbf{Backup}\\
  \hline
  \hline
  1,000      & $8.9 \pm 3$      & $2919.2 \pm 54$    \\
  10,000 & $1.0 \pm 1$     & $493.0 \pm 22$  \\
  100,000    & $0.1 \pm 0.3$        & $52.1 \pm 7$      \\
  1,000,000    & $0.01 \pm 0.1$   & $5.3 \pm 2$ \\
  \hline
\end{tabular}
\caption{Average number of times a user is selected as a committee member or a backup neighbor in one hundred rounds of FL, for a number of user $n\in\{10^3,10^4,10^5,10^6\}$.}
\label{table:simComBack}
\end{center}
\end{table}

\begin{figure}[t]
    \includegraphics[width=1.0\linewidth]{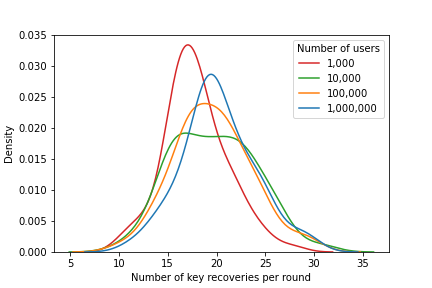}
    \caption{Number of key recoveries per round across 100 secure aggregation rounds. $\gamma=0.2$, $\delta=0.2$}
    \label{fig:nBack}
\end{figure}

Dropout of committee members has a direct impact on the overhead of {\ours}. This is because when committee member $j$ drops out, the server must run a key-recovery protocol with the backup neighbors of $j$.
Figure~\ref{fig:nBack} shows the empirical CDF of the number of key-recoveries necessary across 100 iterations of the secure aggregation protocol. The average number of key-recoveries for $n=10^5$ was $19 \pm 4$, with a maximum of $31$ during one iteration. As shown in Figure~\ref{fig:nBack}, increasing the number of users has little impact on the average number of key recoveries. This is because committee size stays almost constant, even if we increase the number of users.

\begin{table}
\begin{center}
\begin{tabular}{lrr}
  \hline
   &\textbf{Bell}&\textbf{\ours}\\
  \hline
  \hline
  Average      & $25,351$      & $10,085$    \\
  Standard deviation & $1,778$     & $5,780$  \\
  Minimum    & $18,891$        & $3,607$      \\
  Maximum    & $31,482$   & $39,838$ \\
  Per-round average & $254$ & $101$ \\
  \hline
\end{tabular}
\caption{Number of messages sent and received per user over $100$ rounds for $n=10,000$ users.}
\label{table:numMes}
\end{center}
\end{table}

Table~\ref{table:numMes} shows the number of messages sent and received by each user across 100 iterations, for both \ours~ and Bell, in the semi-honest version of both protocols. Here we set $n=10^4$, $m=1$, $\gamma=0.05$ and $\delta=0.33$. The average number of messages sent and received was $10,085\pm 5,780$ over 100 rounds, i.e., the average number of messages sent and received per user per round is $101$. Table~\ref{table:numMes} also shows the number of messages sent and received per users in Bell.
On average, a user in Bell sends and receives $25,351 \pm 1,778$ messages.
The average overhead of a user in Bell is higher that the average user overhead in \ours.

\begin{figure}[t]
    \includegraphics[width=0.9\linewidth]{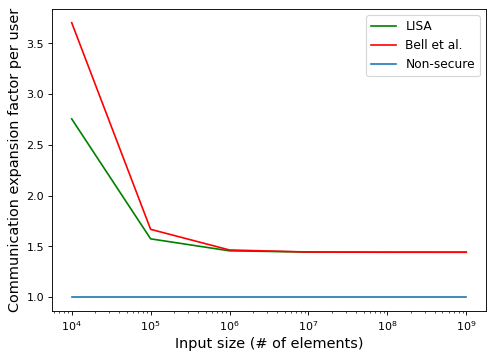}
    \caption{Average communication expansion factor per user per round for increasing input size $m$ in the semi-honest case with $n=100,000$, $\gamma=0.05$, and $\delta=0.33$.}
    \label{fig:bytes}
\end{figure}

In Figure~\ref{fig:bytes} we compare the average communication expansion factor per user during one round of secure aggregation with the one estimated from Bell~\etal~\cite{DBLP:conf/ccs/BellBGL020}. The non-secure protocol, i.e., the one where users send their input in plaintext, is used as baseline. As Figure~\ref{fig:bytes} shows, the bandwidth required by {\ours} is lower than the one needed in~\cite{DBLP:conf/ccs/BellBGL020}, until $m=10^6$. 
For larger input sizes, the expansion factor is similar for both protocols since the number of bytes sent and received per user is very high due to the large input vector. This was also observed in Bonawitz~\etal~ \cite{DBLP:conf/ccs/BonawitzIKMMPRS17}, where the communication expansion factor is small for large inputs. 

\section{Discussion}\label{sec:discussion}

\subsection{Integrity Protection}

\ours{} guarantees input privacy and treats attacks against the integrity of the aggregation as out of scope. There are multiple reasons behind this choice. First, by focusing only on privacy, we allow for a fair comparison with other SA protocols that, similar to \ours, focus on input privacy and treat attacks against integrity as out of scope~\cite{DBLP:conf/ccs/BonawitzIKMMPRS17,DBLP:conf/ccs/BellBGL020,DBLP:journals/iacr/GuoPSBB22,cryptoeprint:2023/486}.
Second, many integrity-protection mechanisms for SA can be added ``on top'' of the SA mechanism. Third, integrity-protection mechanisms for SA have limited guarantees. In particular, integrity-protection mechanisms can ensure that the aggregated value is the sum of a server-chosen set of user inputs. However, no integrity protection mechanism can prevent a malicious user from providing fabricated input to, e.g., bias the output of the computation. Mitigating fabricated inputs requires additional techniques (e.g., range proofs~\cite{DBLP:conf/sp/BunzBBPWM18}) that, however, cannot completely solve the problem. The real benefit that integrity protection mechanisms bring to SA protocols is what Lindell defines as ``independence of inputs''~\cite{lindellmpc}. That is, a malicious user can send fabricated input to the protocol but cannot pick her input after seeing the inputs of honest users (or their sum). In a nutshell, the adversary must commit to malicious inputs before the protocol output is revealed.

In Appendix~\ref{sec:ours+} we show how to add an integrity protection mechanism to \ours. The protocol that guarantees input privacy and integrity is inspired by VeriFL~\cite{DBLP:journals/tifs/GuoLLGHDB21,patchedVeriFL}---the state of the art for SA protocols that guarantee integrity. Note that, shortly after the publication of the original VeriFL paper~\cite{DBLP:journals/tifs/GuoLLGHDB21}, a new version of VeriFL has appeared~\cite{patchedVeriFL}. The new version fixes a flaw in the original protocol that may break input privacy. Further, the new version provides a functionality for the simulation-based proof. In the following, we will use the same functionality to prove the security of our scheme.

VeriFL uses Linearly-Homomorphic Hashing and a commitment scheme to add integrity protection to SA protocols. The downs-side of VeriFl is the linear communication complexity due to the fact that each user must verify the commitments of all other users before accepting the aggregate output as valid. We also use a (linearly-homomorphic) commitment scheme to verify that inputs are actually aggregated in the protocol output. However we take advantage of the committee members to spare each user from verifying the commitments of all other users---thereby avoiding a protocol with linear complexity for regular users.
\smallskip

\noindent\textbf{Additional cryptographic tools.}
We use a commitment scheme $(\CommGen,\CommVfy)$ that is linearly homomorphic. Namely, we denote by~$\commi\gets\CommGen(m,r)$ a commitment to message $m$ with randomness $r$; opening the commitment requires revealing both $m$ and $r$. To verify whether $m,r$ is a valid opening for $\commi$, one can run $\CommVfy(\commi,m,r)$ that provides a $\TRUE/\FALSE$ output. Given $\commi_1\gets\CommGen(m_1,r_1)$ and $\commi_2\gets\CommGen(m_2,r_2)$, the homomorphic property provides a ``multiplication'' operations $\odot$ such that $\commi_1\odot\commi_2 = \CommGen(m_1+m_2,r_1+r_2)$.
In other words, $\CommVfy(\commi_1\odot\commi_2,m_1+m_2,r_1+r_2)$ outputs $\TRUE$.

To prove security against malicious adversaries, we use a commitment scheme as above with the additional properties of equivocability and extractability.
Equivocation means that there exists a trapdoor $td$ and an algorithm $\CommEqv$ allowing the simulator to open a commitment to an arbitrary value. That is, given $\commi\gets\CommGen(m,r)$ and a message $m'$, with $m' \neq m$, the simulator can run $r'\gets\CommEqv(td,\commi,m,r,m')$ such that $\CommVfy(\commi,m',r') = \TRUE$.
Similarly, a commitment scheme is extractable if the simulator can leverage a trapdoor to extract the message from a commitment~$\commi$ by invoking $m\gets \CommExt(td,\commi)$.
We note that commitment schemes with the above properties were proposed in~\cite{DBLP:conf/acns/AbdolmalekiBLS019,DBLP:conf/pkc/CascudoDDGNT15}.
\smallskip

We refer to the new protocol as \ours+. The details are shown in Figure~\ref{fig:protocol:integrity} in Appendix~\ref{sec:ours+}. In a nutshell, we let each user $i$ commit to her input $x_i$ with randomness $r_i$ and send the commitment $\commi_i$ to all of the committee members.
Hence each committee member $j$ can multiply the commitments received from the server so to obtain a commitment to the sum of the inputs, say $C_j=\bigodot_{i\in\users}\commi_i$. Each committee member $j$ will publish $C_j$ and the protocol aborts as soon as one of the published $C_j$ differs from the others. As long as one committee member is honest, we ensure that (i) the server provides the same set of commitments to the committee members, and (ii) that all committee members publish the same  $C_j$; if one of those conditions does not hold, the protocol is aborted. Note that $C_j$ is a commitment to $y=\sum_{i\in\users}x_i$ with randomness $r=\sum_{i\in\users}r_i$.
At the same time, each user $i$ secret-shares the randomness $r_i$ used to create $\commi_i$ across all of the committee members. Hence, committee member $j$ receives $r_{i,j}$ such that $r_i=\sum_{j\in\committee}r_{i,j}$; as long as one committee member is honest, $r_i$ is not leaked to the adversary and commitment $\commi_i$ does not leak $x_i$.

Further, each committee member computes and publishes the sum of received random shares.
That is, committee member $j$ publishes $\rho_j=\sum_{i\in\users}r_{i,j}$. Note that $\sum_{j\in\committee}\rho_j = \sum_{i\in\users}r_i$.
Hence when the server publishes the aggregate sum $y$, a user accepts it as valid only if $\CommVfy(C_j,y,\sum_{j\in\committee}\rho_j)$  outputs $\TRUE$.

\smallskip

The security proof for \ours+ can be found in Appendix~\ref{sec:lisaplus:security}. We also analyze the theoretical overhead of \ours+ in Appendix~\ref{sec:lisaplus:overhead}.

\subsection{Parameters choice}
\label{sec:discussion:parameters}

As in all Byzantine fault-tolerant systems, under-estimating the corruption rate inevitably leads to a security violation. This holds true for all existing secure-aggregation protocols, including {\ours}. Notice, however, that security degrades gracefully: the more corrupt users there are overall (i.e., the higher~$\gamma$), the more likely it is that the number of corrupt committee members exceeds the tolerated threshold, i.e., $\size{\committee \cap \cusers} > \tilde{c}$. For the sake of example, a {\ours} deployment with set-up corruption rate $\gamma=0.1$, $\beta=0.7$, and security parameter $\lambda = 40$,
if the actual corruption rate is $\gamma^*=\gamma + 1\%$,
then the probability of violating security would go from $2^{-40}$ to $2^{-38}$.
A conservative estimation of the corruption rate~$\gamma$ can prevent security violations, however, it has the effect of increasing the committee size. For example, estimating the corruption rate to be $0.25$ instead of $0.10$ increases the committee size from 45 to 71 for 10,000 users and $\delta=0.1$.

Similarly, under-estimating the dropout rate~$\delta$ might prevent the server from unblinding the aggregated input due to too few committee members/backup neighbors being responsive, which would result in wasting a training round. Also in this case, however, the robustness of {\ours} degrades gracefully (for the same reason as above).
To prevent wasting training rounds, here too one could opt for a conservative estimation of~$\delta$, with the effect of increasing the committee size. Here again, estimating the drop-out rate to be $0.25$ instead of $0.10$ increases the committee size from 45 to 71 for 10,000 users and $\gamma=0.1$. 

\section{Conclusions}
\label{sec:conclusions}
In this paper we have presented {\ours}, a LIghtweight Secure Aggregation protocol that leverages a public source of randomness. {\ours} conceals individual user inputs with noise, and leverages public randomness to select a (small) subset of users designated to remove the noise from the sum of individual inputs and reveal the aggregation to the server.
Our protocol completes in 6 rounds but most of the users are required to be online only during the first two rounds. Further, their communication overhead is asymptotically equal to the communication overhead of a non-private protocol where users send their inputs in the clear to the server. We have evaluated {\ours} both theoretically and by means of software prototype. Results show that {\ours} outperforms state-of-the-art secure aggregation protocols in terms of per-user communication complexity.

\bibliographystyle{ACM-Reference-Format}
\bibliography{references}

\appendix
\section{Correctness of {\ours}}
\label{sec:correctness:proof}

In this section we show formally that {\ours} allows the Server to obtain the aggregated sum of the user's inputs despite a limited number of users dropping out during the execution.

Recall that in Round 2, after the committee members and their backup neighbors have been selected, each user~$i$ derives two shared secrets with each committee member~$j$:
a PRG seed~$k^*_{i,j} \gets \KAprf(\textsc{"prg"};\sk_i,\ppk_j)$ to derive the input mask,
and, if $i\in\backupset_j$, a symmetric key~$k^e_{i,j}\gets \KAenc(\textsc{"enc"};\sk_j,\pk_i)$ to retrieve encrypted shares of~$j$'s secret key.
User~$i$ then blinds her input by computing $\ct_i \gets x_i+\sum_{j\in\committee} F(k^*_{i,j})$.
Similarly to regular users, a committee member $j$ masks her input with $k$ blindings, where~$k^*_{j,j} \gets \KAprf(``\textsc{prg}'';\ssk_j,\pk_j)$ is a secret ``shared with herself''.

In Round 3, committee member~$j$ computes the partial blinding as $\partial_j \gets \sum_{i\in\users'_1} F(k^*_{j,i})$, where $k^*_{j,i} = \KAprf(\textsc{"prg"};\ssk_j,\pk_i) = k^*_{i,j}$.

After receiving the blinded inputs from users and the partial blindings from committee members, the server proceeds as follows.
If $\committee_{drop} = \emptyset$, it computes the aggregated input as:
\begin{flalign}
    y &= \sum_{i\in\users'_1} \ct_i - \sum_{j\in\committee} \partial_j\\
      &= \sum_{i\in\users'_1} \left( x_i + \sum_{j\in\committee}F(k^*_{i,j})\right) - \sum_{j\in\committee} \sum_{i\in\users'_1} F(k^*_{i,j})\\
      &= \sum_{i\in\users'_1} x_i
\end{flalign}

In case $\committee_{drop} \neq \emptyset$, the server reconstructs the secret keys of dropped committee members 
using the shares released by the backup neighbors (Rounds~4 and~5), and for each~$j\in\committee_{drop}$ it computes the partial blinding~$\partial_j$ as the committee member would do.
Let $\committee_{alive}=\committee\setminus\committee_{drop}$.
The server computes the aggregate as:
\begin{flalign}
    y &= \sum_{i\in\users'_1} \ct_i - \sum_{j\in\committee_{alive}} \partial_j - \sum_{j\in\committee_{drop}} \partial_j\\
      &= \sum_{i\in\users'_1} \ct_i - \sum_{j\in\committee} \partial_j\\
      &= \sum_{i\in\users'_1} x_i.
\end{flalign}

\section{Security of {\ours}}
\label{sec:simulation:proofs}

To prove security of~{\ours}, we follow the standard simulation-based paradigm and show that the view of any attacker against the protocol can be simulated using only the input of the corrupted parties and the protocol output.
Intuitively, this means that corrupted parties learn nothing more than their own inputs and the intended protocol leakage (i.e., the aggregated input in our case).




We show security for the malicious version of {\ours} (security in the honest-but-curious setting follows directly).
Recall that in the malicious setting, the adversary can corrupt the server and a subset of users~$\cusers\subset \users$ of size $\size{\cusers} = \gamma n$. Simulating the view of such an adversary requires accounting for the intended leakage of the protocol. In particular, {\ours} allow the server to learn the aggregated input of any subset $U\subset \users_1'$ as long as $|U|>\alpha n$. Thus, we give the simulator one-time access to an oracle providing the sum of the inputs by honest users in any subset~$U$ of size at least $\alpha n$.
We note that the first functionality also appears in~\cite{DBLP:conf/ccs/BonawitzIKMMPRS17,DBLP:conf/ccs/BellBGL020}. 
%

For a subset~$U\subseteq \users$, let~$x_i$ denote the input of user~$i\in U$.
Below we use the shorthand $x_{\users\setminus\cusers}$ for the set of inputs~$\{ x_i \}_{i\in\users\setminus\cusers}$.
 The $\alpha\textrm{--}sum$ functionality $\functSUM$ over~$x_{\users\setminus\cusers}$ is parameterized by~$\alpha$, takes as input a subset~$U\subseteq \users\setminus\cusers$, and outputs the aggregation of inputs in~$U$, if $U$ is sufficiently large, else it returns~$\bot$:
\begin{equation}
    \functSUM_{x_{\users\setminus\cusers},\alpha}(U) =
    \begin{cases}
        \sum_{i\in U} x_i & \text{if }\size{U} \geq \alpha \size{\users},\\
        \bot                & \text{otherwise.}
    \end{cases}
\end{equation}

We will provide a security proof in the random oracle model. In particular we model PRG and key derivation functions ($F$ and $\KA$) with a random oracle $\RO\colon\{0,1\}^*\to\{0,1\}^*$ that the simulator will build on the fly.

Given $n,k,t,l$ and a set of corrupted parties $\cusers$, we denote by~$\adv$ the (poly-time) algorithm that defines the strategy of the adversary. This includes logic to decide which messages are sent to and received by corrupted parties and if messages sent by honest parties are ever delivered. Thus, $\adv$ can also decide to make an honest user abort (e.g, by sending her a malformed message) or make an honest user look as if she dropped out (e.g., by not delivering her messages).

Let $\REAL_{n,t,k,l,\gamma,\delta,\cusers}(\adv)$ be the random variable defining the combined view of all corrupted parties during an execution of {\ours}, distributed over the random choices of the honest parties, randomness of the adversary, randomness used to setup cryptographic primitives, and the random oracle. We will now present a simulator $\SIM$ that can emulate the real view of corrupted parties by running $\adv$ and by simulating honest parties on dummy inputs.

We prove the following result.

\begin{theorem}[Privacy against malicious adversaries]
    For every PPT adversary $\adv$,
    all $n,t,k,l \in \NN$ with $k < n$ and $t < l$,
    every set~$\cusers\subset\users$,
    all corruption and dropout rates~$\gamma,\delta \in (0,1)$,
    and all set of inputs~$x_{\users\setminus\cusers}$,
    if the events $S_1$ and $S_2$ hold (cf.~Section~\ref{sec:analysis:security}),
    then there exists a simulator $\SIM$ with access to~$\functSUM$ such that the output of the simulator is computationally indistinguishable from $\REAL_{n,t,k,l,\gamma,\delta,\cusers}(\adv)$.
    In other words:
    \begin{equation*}
    \REAL_{n,t,k,l,\gamma,\delta,\cusers}(\adv,x_{\users\setminus\cusers})
    \equiv
    \SIM^{\functSUM}_{n,t,k,l,\gamma,\delta,\cusers}(\adv)
    \end{equation*}

\end{theorem}

\begin{proof}
We show a sequence of hybrids, starting from the real protocol execution and terminating with a simulated execution that does not use any of the honest parties'inputs.

\begin{description}

    \item[Hyb$_0$]
    The real execution of the protocol.

    \item[Hyb$_1$]
    A simulator who knows the inputs~$\{x_i\}_{i\in\users \setminus \cusers}$ of honest users emulates the real execution by running the protocol with the adversary. This includes simulating the PKI, the randomness beacon, and the random oracle~$\RO$ on the fly.

    \item[Hyb$_2$]
    For every pair of honest users~$i,j$ with $i\in\users$ and $j\in\committee$,
    the simulator replaces the shared symmetric key $k^e_{j,i}\gets\KAenc(\textsc{"enc"};\sk_j,\pk_i)$
    derived by~$j$ in Round~2(\ref{tag:hybrid1:1}) with a uniformly chosen value~$r_{j,i}$ and, to ensure consistency, it also replaces the corresponding key $k^e_{i,j}$ derived by user~$i$ in Round~5(\ref{tag:hybrid1:2}) with the same value~$r_{j,i}$.
    Indistinguishability from the previous hybrid follows from the properties of~$\KA$.

    \item[Hyb$_3$]
    The simulator aborts if, for every pair of honest users~$i,j$ with $i\in\users$ and $j\in\committee$,
    the adversary makes~$i$ deliver a different ciphertexts than the encryption~$E_{j,i}$ generated in Round~2(\ref{tag:hybrid3:1}), and decryption in Round~5(\ref{tag:hybrid3:2}) does not fail.
    Indistinguishability from the previous hybrid follows from the INT-CTXT security of the AE scheme.

    \item[Hyb$_4$]
    For every pair of honest users~$i,j$ with $i\in\users$ and $j\in\committee$,
    the simulator replaces all values $E_{j,i}$ in Round~2(\ref{tag:hybrid3:1}) with encryptions of zero.
    Notice that the simulator still returns the real shares~$S_{j,i}$ to the Server in Round~5(\ref{tag:hybrid4:2}).
    Indistinguishability from the previous hybrid follows from the IND-CPA security of the AE scheme.

    \item[Hyb$_5$]
    The simulator aborts its execution if the adversary forges a valid signature~$\sigma'_i$, on behalf of any honest user~$i$, on a set~$\committee'_{drop}$ different from the set~$\committee_{drop}$ signed by~$i$, and yet the signature verification in Round~5(\ref{tag:hybrid5:1}) succeeds.
    Indistinguishability from the previous hybrid follows from the unforgeability of the digital signature scheme.
    From this hybrid on, we are sure that all honest backup neighbors agree on the (now well-defined) set $\committee_{drop}$ of committee members the server declares to have dropped out.


    \item[Hyb$_6$]
    The simulator aborts if,
    for every pair of honest users~$i,j$ with $i\in\users$ and $j\in\committee$,
    the adversary queries the random oracle~$\RO$ on input~$k^*_{j,i}$ before Round~6.
    Indistinguishability from the previous hybrid follows from the properties of the secret sharing scheme and event~$S_2$ (cf.~Section~\ref{sec:analysis:security}), preventing the adversary from recovering~$\{S_{j,i}\}_{j\in \committee\setminus \cusers}$.

    \item[Hyb$_7$]
    For every pair of honest users~$i,j$ with $i\in\users$ and $j\in\committee$,
    the simulator replaces the shared PRG seeds~$k^*_{i,j}$ and $k^*_{j,i}$, computed by~$i$ as $k^*_{i,j} \gets \KAenc(\textsc{"prg"};\sk_i,\ppk_j)$ in Round~2(\ref{tag:hybrid7:1}), and by~$j$ as~$k^*_{j,i} \gets \KAenc(\textsc{"prg"};\ssk_j,\pk_i)$ in Round~3(\ref{tag:hybrid7:2}), with a value~$r_{i,j}$ chosen uniformly at random,
    and programs the random oracle so that $F(k^*_{i,j}) = F(r_{i,j})$  for every~$j\in\committee_{drop}$; this ensures that the adversary obtains consistent values when computing~$F(k^*_{i,j})$ in Round~6(\ref{tag:hybrid6:1}).
    Indistinguishability from the previous hybrid follows from the key indistinguishability of~$\KA$.

    \item[Hyb$_8$]
    For all honest users~$i\in\users'_1$, the simulator replaces the blinded value~$c_i \gets x_i + \sum_{j\in\committee} F(k^*_{ij})$ with a randomly chosen value, and programs the random oracle (by changing the values of the masks ~$\{F(k^*_{ij})\}_{j\in \committee_{alive}\setminus \cusers}$ each user~$i$ shares with honest, alive committee members) to ensure consistency of the final output, so that:
    \begin{equation}
        \sum_{j\in \committee_{alive}\setminus \cusers} F(k^*_{ij}) = c_i - x_i - \sum_{\substack{j\in  \committee_{drop}\cup \\ (\committee_{alive}\cap \cusers)}}F(k^*_{ij}) .
    \end{equation}
    Notice that this step is possible as long as the set~$\committee_{alive}\setminus \cusers$ of honest and alive committee members is non-empty.
    This condition is guaranteed by event~$S_1$ (cf.~Section~\ref{sec:analysis:security}).

    Concretely, we let the simulator choose a user~$\hat j \in \committee_{alive}\setminus \cusers$ and program the random oracle~$F$ so that:
    \begin{equation}
        F(k^*_{i \hat j}) = c_i - x_i - \sum_{\committee\setminus \{\hat j\}} F(k^*_{ij}) .
    \end{equation}
    Indistinguishability from the previous hybrid follows from the properties of the random oracle.

    \item[Hyb$_9$]
    For every pair of honest users~$i,j$ with $i\in\users$ and $j\in\committee$,
    the simulator programs the random oracle so that every honest users~$i\in\users'_1\setminus \cusers$ computes the mask by replacing the real input~$x_i$ with a randomly chosen value, i.e., instead of
    \begin{equation}
        F(k^*_{ij}) = c_i - x_i - \sum_\committee F(k^*_{ij}) ,
    \end{equation}
     it sets
    \begin{equation}
        F(k^*_{ij}) = c_i - w_i - \sum_\committee F(k^*_{ij}) ,
    \end{equation}
    where all $w_i$'s are chosen at random subject to $\sum_{i\in\users'_1\setminus \cusers} w_i = \sum_{i\in\users'_1\setminus \cusers} x_i$.
    Indistinguishability from the previous hybrid follows from the properties of the random oracle.

    \item[Hyb$_{10}$] Notice that in the last hybrid, the simulator no longer needs the individual inputs~$x_i$ of honest users in~$\users'_1$, it only needs their sum.
    We can therefore let~$\SIM$ emulate the protocol by querying the summation functionality~$\functSUM$, on input the set~$U = \users'_1 \setminus \cusers$, and then choose the values~$w_i$ so that they sum up to the output~$\functSUM(U)$ returned by the functionality.
    This last hybrid is perfectly indistinguishable from the previous hybrid.
    As the simulator does not use the inputs of the honest parties, this concludes the proof.
\end{description}
\end{proof}

\section{\ours+}
\label{sec:ours+}

Figure~\ref{fig:protocol:integrity} provides the details of \ours+.

\begin{figure*}[t]
\raggedright

\textbf{Parties:} Server and users~$\users = [n]$.

\textbf{Public parameters:} input domain~$\XX$, fraction of drop-outs~$\delta$, fraction of corruptions~$\gamma$, security parameter for cryptographic primitives $\sep$, committee size~$k$, backup-neighborhood size~$\ell$, secret sharing reconstruction threshold~$t$, minimum fraction of aggregated inputs~$\alpha$, maximum number of corrupt committee members~$\tilde{c}$.

\textbf{Prerequisites:} Each user $i\in\users$ has key-pairs computed as $(\sk_i,\pk_i) \gets \KAgen(1^\sep)$, $(\ssk_i,\ppk_i) \gets \KAgen(1^\sep)$ and $(\Ssk_i,\Spk_i)\gets\DSgen(1^\sep)$; public keys $(\pk_i,\ppk_i,\Spk_i)$ are registered with the PKI.

For~$\users$ and~$r \in [6]$ and, we denote by~$\users_r$ the set of users that complete the execution of round~$r$ without dropping out, and we denote by~$\users_r '$ the set of users the server knows have completed round~$r$. It holds $\users'_r \subseteq \users_r \subseteq \users_{r-1}$ for all~$r\in [6]$.

\smallskip
\scalebox{0.8}{%
\begin{minipage}{.6\textwidth}
\textbf{Round 1}
\begin{description}[resume]
    \item{Each party}
        \begin{enumerate}\setlength{\itemindent}{-.3in}
        \item receives random seed~$Q$
        \item selects committee $\committee\gets\nchoosek(Q,\users,k)$
        \end{enumerate}
\end{description}
\textbf{Round 2}
\begin{description}[resume]
    \item{Committee member~$j \in \committee$}
        \begin{enumerate}\setlength{\itemindent}{-.3in}
        \item selects backup neighbors $\backupset_j\gets\nchoosek(Q||j,\users\setminus\{j\},\ell)$
        \item fetches public keys of backup neighbors $\{\pk_i\}_{i\in \backupset_j}$ from the PKI
        \item derives symmetric keys: $\{k^{e}_{j,i} \gets \KAenc(\textsc{"enc"};\sk_j,\pk_i)\}_{i\in \backupset_j}$
        \item secret shares key~$\ssk_j$: $\{S_{j,i}\}_{i\in\backupset_j}\gets\SSshare(\ell;t;\ssk_j)$
        \item encrypts shares of $\ssk_j$: $\{E_{j,i}\gets\AEenc(k^e_{j,i},S_{j,i})\}_{i\in\backupset_j}$\label{tag:INT:hybrid3:1}
        \item sends $\{(j;i;E_{j,i})\}_{i\in\backupset_j}$  to the server
        \end{enumerate}
    \item{User~$i\in\users$}
        \begin{enumerate}\setlength{\itemindent}{-.3in}
        \item fetches public keys $\{\ppk_j\}_{j\in\committee}$ from the PKI
        \item derives symmetric keys:~$\{k^{*}_{i,j} \gets \KAprf(\textsc{"prg"};\sk_i,\ppk_j)\}_{j\in\committee}$
        \item computes blinded input: $\ct_i \gets x_i+\sum_{j\in\committee}F(k^{*}_{i,j})$
        \item sends~$\ct_i$ to the server 
        {\color{blue}
        \item computes commitment to $x_i$: $\commi_i\gets\CommGen(x_i,r_i)$ where $r_i$ is a random value
        \item computes signature on $\commi_i$: $\sigma_i\gets \DSsign(\Ssk_i;\commi_i)$
        \item computes k-out-of-k shares of $r_i$: $\{r_{i,j}\}_{j\in\committee}\gets \SSshare(k,k;r_i)$
        \item derives symmetric keys: $\{k^{s}_{i,j} \gets \KAenc(\textsc{"enc"};\sk_i,\ppk_j)\}_{j\in \committee}$
        \item encrypts the random shares of $r_i$: $\{E^s_{i,j}\gets\AEenc(k^s_{i,j},r_{i,j})\}_{j\in\committee}$
        \item sends $\{(i;j;E^s_{i,j})\}_{j\in\committee}, (\commi_i,\sigma_i)$  to the server
        }
        \end{enumerate}
\end{description}

\textbf{Round 3}
\begin{description}
    \item{Server}
        \begin{enumerate}\setlength{\itemindent}{-.3in}
        \item receives encrypted key shares~$\{(j;i;E_{j,i})\}_{i\in \backupset_j}$  from $j\in\committee\cap\users'_2$ \\ and sends each of them to corresponding backup neighbor
        \item receive blinded inputs~$\ct_i$ from users~$i\in\users_2 '$
        \item aggregates input $c_{agg}\gets\sum_{i\in\users_2'}c_i$
        \item sends~$\users'_2$ to committee members $j\in \committee\cap\users'_2$.
        {\color{blue}
        \item receives encrypted random shares $\{(i;j;E^s_{i,j})\}_{j\in\committee}$ from $i\in\users'_2$ \\ and sends each of them to the corresponding committee member
        \item receives commitments and signatures $\{(\commi_i,\sigma_i)\}_{i\in\users'_2}$ and send all of them to all committee members
        }
        \end{enumerate}
    \item{Committee member $j\in\committee\cap\users'_2$}
        \begin{enumerate}\setlength{\itemindent}{-.3in}
          \item receives $\users'_2$ from the server
          \item if $\size{\users'_2}<\alpha n$, aborts
          {\color{blue}
          \item receives commitments and signatures $\{(\commi_i,\sigma_i)\}_{i\in\users'_2}$
          \item asserts that signatures ar valid: \\
                if $\size{\{i\in\users'_2:\quad \DSverif(\Spk_i;\sigma_i;\commi_i) = \TRUE\}}<\size{\users'_2}$ then aborts
          \item aggregates the commitments: $C_j\gets\bigodot_{i\in\users'_2}\commi_i$ \label{tag:INT:hybrid:10:1}
          \item sign the aggregate commitment and set $\users'_2$: \\ $\sigma^C_j\gets\DSsign(\Ssk_j;C_j||\users'_2)$
          \item receive encrypted random shares: $\{(i;j;E^s_{i,j})\}_{i\in\users'_2}$
          \item derives symmetric keys~$\{k^{s}_{j,i} \gets \KAprf(\textsc{"enc"};\ssk_j,\pk_i)\}_{i\in\users'_2}$
          \item decrypt random shares $\{r_{i,j}\gets\AEdec(k^s_{j,i};E_{i,j})\}_{i\in\users'_2}$
          \item computes the sum or random shares $\rho_j\gets\sum_{i\in\users'_2}r_{i,j}$
          \item signs the sum or random shares $\sigma^\rho_j\gets \DSsign(\Ssk_j;\rho_j)$
          \item sends $\rho_j, \sigma^\rho_j, C_j, \sigma^C_j$ to the server
          }
          \item fetches public keys $\{\pk_i\}_{i\in\users'_2}$ from the PKI
          \item derives symmetric keys~$\{k^{*}_{j,i} \gets \KAprf(\textsc{"prg"};\ssk_j,\pk_i)\}_{i\in\users'_2}$\label{tag:INT:hybrid7:2}
          \item computes partial blinding $\partial_j\gets\sum_{i\in\users'_1}F(k^{*}_{j,i})$
          \item sends~$\partial_j$ to the server
        \end{enumerate}
\end{description}

\end{minipage}\quad\quad

\begin{minipage}{.65\textwidth}

\textbf{Round 4}

\begin{description}
        \item{Server} \\
        \texttt{//let $\committee_{alive}:=\committee\cap \users'_3$ and $\committee_{drop} :=\committee \setminus \committee_{alive}$ and $\backupset := \bigcup_{j\in\committee} \backupset_j$}
        \begin{enumerate}\setlength{\itemindent}{-.3in}
        {\color{blue}\item receives $\{(\rho_j, \sigma^\rho_j, C_j, \sigma^C_j)\}_{j\in\committee_{alive}}$ from committee members in~$\committee_{alive}$}
        \item receives $\{\partial_j\}_{j\in\committee_{alive}}$  from committee members in~$\committee_{alive}$
        \item if $\size{\committee_{alive}} = k$ jumps to step~(\ref{line:M:final:decryption}) of Round 6 
        \item sends~$\committee_{drop}$ to all users in $\backupset$\
        \end{enumerate}\setlength{\itemindent}{-.3in}
   \item{Backup neighbor~$i\in \backupset$}
        \begin{enumerate}\setlength{\itemindent}{-.3in}
        \item receives~$\committee_{drop}$ from the server
        \item if $\size{\committee_{drop}} \geq k - \tilde{c}$ then aborts 
        \item computes signature $\sigma_i\gets\DSsign(\Ssk_i;\committee_{drop})$
        \item sends $\sigma_i$ to the server
        \end{enumerate}
\end{description}

\textbf{Round 5}
\begin{description}
    \item{Server}
        \begin{enumerate}\setlength{\itemindent}{-.3in}
            \item receives $\{\sigma_i\}_{i\in \backupset \cap \users'_3}$ from users $\backupset \cap \users'_3$\\
                and forwards them to all users in $\backupset \cap \users'_3$
        \end{enumerate}

    \item{Backup neighbor~$i\in \backupset \cap \users'_3$}
        \begin{enumerate}\setlength{\itemindent}{-.3in}
        \item receives $\{\sigma_i\}_{i\in \backupset \cap \users'_3}$
        \item fetches $\{\Spk_i\}_{i\in \backupset \cap \users'_3}$ from the PKI
        \item computes $\backupset_{ack} \gets \{ l\in\backupset\cap \users'_4 : \DSverif(\Spk_l;\sigma_l;\committee_{drop}) = \TRUE\}$ 
        \item if $\size{\backupset_j\cap\backupset_{ack}} < t$ for any~$j\in\committee$ then aborts
        \item for any $j\in\committee_{drop}$ such that $i\in\backupset_j$,
            \begin{enumerate}\setlength{\itemindent}{-.3in}
            \item fetches $\pk_j$ from the PKI
            \item derives symmetric key~$(k^{e}_{i,j}) \gets \KAenc(\textsc{"enc"};\sk_i,\pk_j)$ 
            \item decrypts $S_{j,i} \gets \AEdec(k^e_{i,j};E_{j,i})$ 
            \item sends $S_{j,i}$ to the server 
            \end{enumerate}
        \end{enumerate}
\end{description}

\textbf{Round 6}
\begin{description}
    \item{Server}
        \begin{enumerate}\setlength{\itemindent}{-.3in}
            \item For each $j\in\committee_{drop}$
            \begin{enumerate}\setlength{\itemindent}{-.3in}
                \item collects shares $\{S_{j,i}\}_{i\in\backupset_j\cap \users'_5}$ and aborts if receives less than $t$ shares
                \item reconstructs secret key $\ssk_j\gets\SSrecon(\{S_{j,i}\}_{i\in\backupset_i\cap \users'_5})$
                \item derives symmetric keys $\{k^{*}_{j,i}\gets \KAprf(\textsc{"prg"};\ssk_j;\pk_i)\}_{i\in\users_2'}$ 
                \item computes missing partial blinding~$\partial'_j\gets\sum_{i\in\users'_2}F(k^{*}_{i,j})$
                {\color{blue}
                \item derives symmetric keys $\{k^{s}_{j,i}\gets \KAprf(\textsc{"enc"};\ssk_j;\pk_i)\}_{i\in\users_2'}$
                \item decrypt random shares $\{r_{i,j}\gets\AEdec(k^s_{j,i};E_{i,j})\}_{i\in\users'_2}$ \label{tag:INT:hybrid:opening:shares:2}
                \item computes the sum or random shares $\rho_j\gets\sum_{i\in\users'_2}r_{i,j}$
                }
            \end{enumerate}
                \item given $\{\partial_j\}_{j\in\committee}=\{\partial_j\}_{j\in\committee_{alive}} \cup \{\partial_j\}_{j\in\committee_{drop}}$, computes the output:\\ 
                    $y\gets c_{agg} - \sum_{j\in\committee}\partial_j$
                {\color{blue}
                \item publishes $y$ and $\users'_2$ and $\{(\rho_j, \sigma^\rho_j, C_j, \sigma^C_j)\}_{j\in\committee_{alive}}$ and $\{\rho_j\}_{j\in\committee_{drop}}$
                }
        \end{enumerate}
    {\color{blue}
    \item{User $i$}
        \begin{enumerate}\setlength{\itemindent}{-.3in}
            \item fetches $\{\Spk_j\}_{j\in\committee_{alive}}$ from the PKI
            \item asserts the following:
            \begin{enumerate}\setlength{\itemindent}{-.3in}
                \item $\size{\committee_{alive}} \geq k - \tilde{c}$
                \item for all pair of indices $j_1,j_2\in\committee_{alive}$:\quad $C_{j_1}=C_{j_2}$
                \item $\DSverif(\Spk_j;\sigma^\rho_j;\rho_j) = \TRUE\ $ for $j\in\committee{alive}$ \label{tag:INT:hybrid:signature:verif:aggopening}
                \item $\DSverif(\Spk_j;\sigma^C_j;C_j||\users'_2) = \TRUE\ $ for $j\in\committee{alive}$ \label{tag:INT:hybrid:signature:verif:aggcomm}
                \item $\CommVfy(C_j,y,\sum_{j\in\committee}\rho_j)=\TRUE$
            \end{enumerate}
        \end{enumerate}
    }

\end{description}

\end{minipage}%
}

\caption{\ours+ secure aggregation protocol with integrity protection. Additional steps compared to Figure~\ref{fig:protocol:both} are {\color{blue} highlighted in blue.}}
\label{fig:protocol:integrity}
\end{figure*}

\section{Theoretical evaluation of \ours+}
\label{sec:lisaplus:overhead}

We theoretically evaluate the communication of users and the server for \ours+.

\paragraph{Regular users: $\asim (km)$ communication.} Regular users --- those not selected as committee members nor as backup neighbors--- are active only during the first two rounds, and the last round to verify the integrity of the computation. They fetch public keys of $k$ committee members from the PKI and send $m$ noisy inputs. They send $k$ shares and $m$ commitments to the server, and receive $km$ aggregated commitments.


\paragraph{Committee members: $\asim (nm+ +\ell)$ communication.} Each committee member fetches $\ell$ public keys from the PKI and sends $\ell$ ciphertexts Round 2. Next, in Round 3, receives up to $n$ public keys and sends $m$ random values. They receive $nm$ commitments, $n$ signatures and shares, and send $m$ aggregate commitment, sum of shares and a pair of signatures.


\paragraph{Backup neighbor: $\asim (k\ell)$ communication.} A user can be selected as a backup neighbor by multiple committee members. Hence, it receives up to $k$ ciphertexts during Round 3. In Round 4, each backup neighbor receives one messages from the server and sends one signature. Next, during Round 5 it receives up to $k\ell$ signatures and sends up to $k$ shares.


\paragraph{Server: $\asim (nm+k^2\ell^2+nk)$ communication.} The server receives $k\ell$ shares and $nm$ blinded inputs during Round 3, and sends $k\ell$ shares. It receives $n$ sets of $k$ shares and $n$ signatures. During Round 4, the server receives up to $km$ partial blindings and sends one message to up to $k\ell$ users. It receives $k$ signatures and aggregated commitments and shares from the committee members. During Round 5, it receives up to $k\ell$ signatures and sends the same number of signatures to each member of each backup neighborhood. Finally, during round 6 it receives up to $k\ell$ shares, and sends $k$ signatures and aggregated commitments and shares to all $n$ users.

We conclude that adding integrity to \ours~ does not have an impact on the asymptotic communication overhead of the regular users. Similarly, VeriFL does not increase the communication overhead of the protocol proposed by Bonawitz~\etal \cite{DBLP:conf/ccs/BonawitzIKMMPRS17}.

We summarize in Table \ref{table:integrity_comparison} the communication overhead with and without integrity protection for \ours and Bonawitz~\etal \cite{DBLP:conf/ccs/BonawitzIKMMPRS17}. Bell~\etal~ do not present a protocol with integrity protection.

\begin{table}
\begin{center}
\begin{tabular}{llp{3cm}}
  \hline
  &\textbf{Communication}&\textbf{Security guarantees}\\
  \hline
  \hline
  \multirow{3}{*}{{\ours}}                   & Regular: $O(m)$           & \multirow{3}{*}{\parbox{3cm}{Input privacy}}\\
    & Committee: $\asim(n+m)$              &                                                         \\
    & Backups:   $\asim(1)$              &                                                          \\
    \midrule
    \multirow{3}{*}{{\ours}+}                   & Regular: $O(m)$           & \multirow{3}{*}{\parbox{3cm}{Input privacy + computation integrity}}\\
    & Committee: $\asim(nm)$              &                                                          \\
    & Backups:   $\asim(1)$              &                                                            \\
    \midrule
  \multirow{2}{*}{VeriFL \cite{patchedVeriFL}} & \multirow{2}{*}{$\asim(n+m)$ }     & \multirow{2}{*}{\parbox{3cm}{Input privacy + computation integrity}} \\
    &  &  \\
  \hline
\end{tabular}
\caption{Theoretical communication overhead.}
\label{table:integrity_comparison}
\end{center}
\end{table} 
\section{Security of \ours+}
\label{sec:lisaplus:security}

For capturing integrity properties of a secure aggregation protocol, we consider the ideal functionality~$\functSUMVERIF$ proposed in the fixed version of the VeriFL protocol~\cite{patchedVeriFL}. Notice that for efficiency purposes, VeriFL aggregates user inputs in batches of size~$\ell$. Our protocol does not require batched aggregation, therefore we here consider $\ell = 1$.
In a nutshell, $\functSUMVERIF$ allows each user to send her input and the adversary to decide the set of user inputs to be aggregated (i.e., this is $\users'_2$ in \ours+). Given the true aggregate $a$ for the set of user inputs decided by the adversary, the functionality allows the adversary to specify which aggregate to be sent to each user. If the adversary asks the functionality to send aggregate $a'\neq a$ to user $i$, then the functionality marks that user as ``cheated'' and sends $a'$ to user $i$. Finally each user can ask the functionality whether it has been cheated. We refer to~\cite{patchedVeriFL} for the details of~ $\functSUMVERIF$.

Let $\REAL_{n,t,k,l,\gamma,\delta,\cusers}(\adv)$ be the random variable defining the combined view of all corrupted parties during an execution of {\ours+}, distributed over the random choices of the honest parties, randomness of the adversary, randomness used to setup cryptographic primitives, and the random oracle. We present a simulator $\SIM$ that emulates the real view of corrupted parties by running adversary $\adv$ and by simulating honest parties on dummy inputs.
We prove the following result:

\begin{theorem}[Privacy and integrity against malicious adversaries]
    For every PPT adversary $\adv$,
    all $n,t,k,l \in \NN$ with $k < n$ and $t < l$,
    every set~$\cusers\subset\users$,
    all corruption and dropout rates~$\gamma,\delta \in (0,1)$,
    and all set of inputs~$x_{\users\setminus\cusers}$,
    if the events $S_1$ and $S_2$ hold (cf.~Section~\ref{sec:analysis:security}),
    then there exists a simulator $\SIM$ with access to~$\functSUMVERIF$ such that the output of the simulator is computationally indistinguishable from $\REAL_{n,t,k,l,\gamma,\delta,\cusers}(\adv)$.
    In other words:
    \begin{equation*}
    \REAL_{n,t,k,l,\gamma,\delta,\cusers}(\adv,x_{\users\setminus\cusers})
    \equiv
    \SIM^{\functSUMVERIF}_{n,t,k,l,\gamma,\delta,\cusers}(\adv)
    \end{equation*}
\end{theorem}

\begin{proof}
    We show a sequence of hybrids, starting from the real protocol execution and terminating with a simulated execution that does not use any of the honest party inputs. As many of these hybrids are the same as in the proof of \ours, we will refer to the proof in Appendix~\ref{sec:simulation:proofs} as ``privacy proof''.

    \begin{description}


    \item[Hyb$_0$--Hyb$_9$:]
    Same as in the privacy proof.

    \item[Hyb$_{10}$]
    The simulator aborts its execution if the adversary forges a valid signature~$\sigma_i$, on behalf of any honest user~$i$, on a commitment~$\commi_i$, and the signature verification in Round~3(\ref{tag:INT:hybrid:10:1}) succeeds.
    Indistinguishability from the previous hybrid follows from the unforgeability of the digital signature scheme.
    This ensures that the adversary cannot manipulate the commitments of honest users.

    \item[Hyb$_{11}$]
    The simulator aborts its execution if the adversary forges a valid signature~$\sigma^\C_j$, on behalf of an honest committee member~$j$, on message~$C_j||\users'_2$, and the signature verification in Round~6(\ref{tag:INT:hybrid:signature:verif:aggcomm}) succeeds.
    Indistinguishability from the previous hybrid follows from the unforgeability of the digital signature scheme.
    From this hybrid on, honest committee members agree on the (now well-defined) set $\users'_2$ of users the server declares to have contributed the aggregation, as well as on the commitments~$C_j$ of the aggregated inputs.

    \item[Hyb$_{12}$]
    For all pairs of honest users~$i\in\users$ and~$j\in\committee$, the simulator replaces key~$k^s_{i,j}$ with a random value. Because of the key indistinguishability of~$\KA$, this hybrid is indistinguishable from the previous one.

    \item[Hyb$_{13}$]
    For all pairs of honest users~$i\in\users$ and~$j\in\committee$, the simulator aborts if the adversary delivers a different ciphertext~$E^s_{i,j}$ than the one generated in Round~2(\ref{tag:INT:hybrid3:1}), yet decryption succeeds.
    Because of the INT-CTXT security of the encryption scheme, the simulator aborts only with negligible probability, hence this hybrid is indistinguishable from the previous one.

    \item[Hyb$_{14}$]
    For all pairs of honest users~$i\in\users$ and~$j\in\committee$, the simulator replaces the encrypted shares~$E^s_{i,j}$ with encryptions of zero. Note however that the simulator still returns the real shares in Round 6(\ref{tag:INT:hybrid:opening:shares:2}).
    Indistinguishability from the previous hybrid follows from the IND-CPA security of the encryption scheme.

    \item[Hyb$_{15}$]
    The simulator aborts its execution if the adversary forges a valid signature~$\sigma^\rho_j$, on behalf of an honest committee member~$j$, on the sum of shares~$\rho_j$, and yet the signature verification in Round 6(\ref{tag:INT:hybrid:signature:verif:aggopening}) succeeds.

    \item[Hyb$_{16}$]
    For every honest user~$i$, the simulator replaces the commitment $\commi_i \gets \CommGen(x_i,r_i)$ with a commitment to a random value~$u_i$.
    To ensure indistinguishability from the previous hybrid, the simulator will now make sure that the aggregated commitments $C_j = \bigodot_{i\in\users'_2}\commi_i$ can be opened to the final aggregation.
    To this end, when the server declares set $\users'_2$, the simulator requests the corresponding aggregation by sending $(\mathsf{ready},\users'_2)$ to $\functSUMVERIF$, and obtains the aggregated value~$y = \sum_{i\in \users'_2} x_i$.
    The simulator further invokes the extraction algorithm $\CommExt$ to extract the inputs of the corrupt users in~$\users'_2$, obtaining $x_i\gets \CommExt(td,\commi_i)$ for all~$i\in \cusers \cap \users'_2$.
    The simulator hence reconstructs the partial sum contributed by the corrupt parties, i.e., $y_\cusers \gets \sum_{i \in \users'_2 \cap \cusers} x_i$, and derives the sum of the honest users'inputs as $y - y_\cusers$.
    To ensure each commitment $C_j= \bigodot_{i\in\users'_2}\commi_i$, computed by committee member~$j$, will open to the aggregated value~$y$, the simulator finally uses the equivocation algorithm to compute openings $r'_i$ and secret-shares $r'_i$ instead of $r_i$ among committee members. Given the security of the secret-sharing scheme, this hybrid is indistinguishable from the previous one.
    \item[Hyb$_{17}$]
    For all honest users~$i\in\users'_2$, the simulator replaces the blinded value~$c_i \gets x_i + \sum_{j\in\committee} F(k^*_{ij})$ with a randomly chosen value, and programs the random oracle (by changing the values of the masks ~$\{F(k^*_{ij})\}_{j\in \committee_{alive}\setminus \cusers}$ each user~$i$ shares with honest, alive committee members) to ensure consistency of the final output, so that:
    \begin{equation}
        \sum_{j\in \committee_{alive}\setminus \cusers} F(k^*_{ij}) = c_i - x_i - \sum_{\substack{j\in  \committee_{drop}\cup \\ (\committee_{alive}\cap \cusers)}}F(k^*_{ij}) .
    \end{equation}
    Notice that this step is possible as long as the set~$\committee_{alive}\setminus \cusers$ of honest and alive committee members is non-empty.
    This condition is guaranteed by event~$S_1$ (cf.~Section~\ref{sec:analysis:security}).

    Concretely, we let the simulator choose a user~$\hat j \in \committee_{alive}\setminus \cusers$ and program the random oracle~$F$ so that:
    \begin{equation}
        F(k^*_{i \hat j}) = c_i - x_i - \sum_{\committee\setminus \{\hat j\}} F(k^*_{ij}) .
    \end{equation}
    Indistinguishability from the previous hybrid follows from the properties of the random oracle.

    \item[Hyb$_{18}$]
    For every pair of honest users~$i,j$ with $i\in\users$ and $j\in\committee$,
    the simulator programs the random oracle so that every honest users~$i\in\users'_2\setminus \cusers$ computes the mask by replacing the real input~$x_i$ with a randomly chosen value, i.e., instead of
    \begin{equation}
        F(k^*_{ij}) = c_i - x_i - \sum_\committee F(k^*_{ij}) ,
    \end{equation}
     it sets
    \begin{equation}
        F(k^*_{ij}) = c_i - w_i - \sum_\committee F(k^*_{ij}) ,
    \end{equation}
    where all $w_i$'s are chosen at random subject to $\sum_{i\in\users'_2\setminus \cusers} w_i y - y_\cusers$.
    Indistinguishability from the previous hybrid follows from the properties of the random oracle.


    \item[Hyb$_{19}$] Notice that in the last hybrid, the simulator no longer needs the individual inputs~$x_i$ of honest users in~$\users'_2$, it only needs their sum which can be obtained by querying $\functSUMVERIF$ on input $(\mathsf{ready}, \users'_2)$ and subtracting from the response~$y$ the partial sum of the corrupt parties $y_\cusers$. 
    As the simulator does not use the inputs of the honest parties, this concludes the proof.

    \end{description}

\end{proof} 

\end{document}